\pdfoutput=1
\documentclass[reprint, twocolumn, preprintnumbers, superscriptaddress, aps, pra]{revtex4-2}
\usepackage{graphicx} 
\usepackage{amsmath, amssymb}
\usepackage{mathtools}
\usepackage{bm} 

\usepackage[caption=false]{subfig}
\usepackage{multirow} 
\usepackage{xcolor} 
\usepackage{parskip} 
\usepackage{enumitem} 
\usepackage{microtype}
\usepackage{booktabs}

\usepackage[T1]{fontenc} 
\usepackage[utf8]{inputenc} 
\usepackage{lmodern} 
\usepackage{orcidlink}

\usepackage{braket} 

\usepackage{tcolorbox} 

\usepackage{hyperref}
\hypersetup{
    colorlinks=true,
    linkcolor=blue,
    citecolor=blue,
    urlcolor=blue
}
\begin{document}

\title{Generative Adversarial Networks for Resource State Generation}
\author{Shahbaz Shaik}
\email{shahbaz.shaik@research.iiit.ac.in}
\affiliation{Centre for Quantum Science and Technology, International Institute of Information Technology, Hyderabad}
\affiliation{Center for Security Theory and Algorithmic Research, International Institute of Information Technology, Hyderabad}

\author{Sourav Chatterjee\,\orcidlink{0000-0002-1920-7365}}
\email{sourav.chat@tcs.com}
\affiliation{TCS Research, TATA Consultancy Services, India}

\author{Sayantan Pramanik}
\email{sayantan.pramanik@tcs.com}
\affiliation{TCS Research, TATA Consultancy Services, India}

\author{Indranil Chakrabarty\,\orcidlink{0009-0001-0415-0431}}%
\email{indranil.chakrabarty@iiit.ac.in}
\affiliation{Centre for Quantum Science and Technology, International Institute of Information Technology, Hyderabad}
\affiliation{Center for Security Theory and Algorithmic Research, International Institute of Information Technology, Hyderabad}

\begin{abstract}

We introduce a physics-informed Generative Adversarial Network framework that recasts quantum resource-state generation as an inverse-design task. By embedding task-specific utility functions into training, the model learns to generate valid two-qubit states optimized for teleportation and entanglement broadcasting. Comparing decomposition-based and direct-generation architectures reveals that structural enforcement of Hermiticity, trace-one, and positivity yields higher fidelity and training stability than loss-only approaches. The framework reproduces theoretical resource boundaries for Werner-like and Bell-diagonal states with fidelities exceeding~98\%, establishing adversarial learning as a lightweight yet effective method for constraint-driven quantum-state discovery. This approach provides a scalable foundation for automated design of tailored quantum resources for information-processing applications, herein exemplified with teleportation and broadcasting of entanglement, and also opens up the possibility of using them in efficient quantum network design.
\end{abstract}

\maketitle

\section{Introduction}

Resource generation is a central challenge across the physical sciences. In classical contexts, it appears as the design of systems with prescribed responses---from photonic crystals that mold the flow of light~\cite{joannopoulos2008photonic} to metamaterials exhibiting negative refractive index~\cite{smith2004metamaterials}. Traditional parametric and variational methods become rapidly intractable as the number of degrees of freedom grows~\cite{molesky2018inverse}. Machine learning (ML) has therefore emerged as a natural paradigm for inverse design, with neural networks (NNs) enabling automated discovery of structures in optics~\cite{malkiel2018plasmonic}, acoustics~\cite{zhao2019acoustic}, and materials science~\cite{schmidt2019recent}. More generally, NNs are being widely employed across various areas of  physics~\cite{carleo2019machine, shlomi2020graph, aragon2019classifying, carrasquilla2020machine, schmidt2019recent, iten2020discovering} for classification tasks~\cite{zhang2002neural}, where the objective is to assign labels to data samples, and for generative modeling~\cite{kingma2013auto, goodfellow2014generative}, where new data are produced after learning underlying data distributions~\cite{Ahmed2021qstcgan}. Generative Adversarial Networks (GANs)~\cite{Goodfellow2020gan} offer an appealing framework for the latter purpose.

In quantum information science, many general challenges in characterizing and controlling quantum systems can be reformulated as machine-learning tasks. For instance, NN–based methods have been successfully applied to address foundational questions~\cite{bharti2020machine}, phase transition detection\,\cite{rem2019identifying}, entanglement\,\cite{ma2018transforming, harney2020entanglement} and nonclassicality\,\cite{sentis2015quantum, PhysRevResearch.2.023150} identification, quantum experiment design\,\cite{krenn2020computer}, error correction\,\cite{torlai2017neural, fosel2018reinforcement}, and device calibration\,\cite{wittler2021integrated}. In particular, the inverse-design viewpoint (mentioned above) applies to quantum states themselves, which act as operational resources for information-processing tasks~\cite{Chitambar2019}. A valid quantum state must satisfy Hermiticity, unit trace, and positive semidefiniteness\,\cite{nielsen2010quantum}, and its resourcefulness is often defined by protocol-dependent criteria such as teleportation fidelity~\cite{horodecki1996teleportation} or broadcasting capability~\cite{chatterjee2016broadcasting}. The analytical characterization of all states fulfilling a given criterion is rarely feasible, especially beyond the two-qubit regime~\cite{horodecki2009quantum}. This makes the systematic discovery of useful resource states a natural candidate for data-driven approaches that can explore and learn such constrained manifolds directly~\cite{Ahmed2021qstcgan}.

Quantum information processing often relies critically on entanglement as a resource. However, this resource can at times be task specific~\cite{ganguly2011entanglement,vempati2021witnessing} and sometimes beyond states and channel, i.e., in network where it can be topology specific ~\cite{mylavarapu2024teleportation}. It can be activated under global unitary operation~\cite{patro2017non} and can be conserved or destroyed under the action of channels~\cite{srinidhi2024quantum,vempati2022unital}. Quantum teleportation is a protocol that exploits shared entanglement and classical communication to transfer an unknown quantum state from one party to another~\cite{bennett1993teleporting}. Maximally entangled Bell states enable perfect teleportation~\cite{bennett1993teleporting,horodecki1996teleportation}. However, there can be other entangled states that can act as a resource in offering the quantum advantage in teleportation (i.e., surpassing its classical limits)~\cite{horodecki1996teleportation}. On the other hand, broadcasting of quantum entanglement is a method to quantify the range of input parameters of a shared state under which it can act as a resource by broadcasting its quantum entanglement into multiple weaker entangled states~\cite{buzek1997broadcasting,chatterjee2016broadcasting}. However, a more general resource theoretic framework was lately used to understand different types of resources including distributed quantum computation~\cite{wehner2018quantum}, superdense coding~\cite{bennett1992communication}, 
distillation~\cite{horodecki2001distillation} and many more~\cite{singh2024controlled, vempati2021witnessing}. The ability to identify and generate these resource states 
is hence essential.

Originally developed to learn complex data distributions, GANs have since been applied across physics, from phase-transition discovery~\cite{hu2017discovering} to particle-collision modeling~\cite{paganini2018accelerating}. Quantum variants (qGANs)~\cite{Lloyd2018qgal,Hu2019qgan} and classical-quantum hybrids~\cite{Ahmed2021qstcgan,Koutn2022neuralnetworkqst} have demonstrated their ability to reproduce valid quantum states, yet targeted generation of resource states optimized for specific tasks remains largely unexplored. Compared with likelihood-based or diffusion models~\cite{ho2020diffusion}, adversarial learning provides a lightweight mechanism in which the discriminator naturally enforces physical validity, making GANs particularly suited for low-dimensional, constraint-dominated problems such as synthesis of density matrices representing quantum states\,\cite{nielsen2010quantum}.

In summary, resource-state generation is a unifying inverse-design problem across physics, from creating metamaterials~\cite{smith2004metamaterials} with target electromagnetic responses~\cite{molesky2018inverse} to engineering molecular structures~\cite{sanchez2018inverse}. In quantum information science, this challenge is amplified: entangled states enable core protocols such as teleportation and broadcasting, yet analytical characterization of useful non-maximally entangled states remains elusive~\cite{Chitambar2019}. In this work, we present a GAN framework that treats quantum resource-state generation as an inverse-design problem. By embedding protocol-specific utility functions directly into training, our model learns distributions of valid two-qubit states acting as a resource for teleportation and broadcasting. We compare three generator architectures---two enforcing physical constraints through decomposition and one unconstrained model trained via loss penalties---and show that structural enforcement yields higher fidelity and training stability. Our results demonstrate adversarial learning as a practical and extensible method for generating tailored quantum resources. Generated quantum resources advantageous for communication can be used for large scale quantum network design~\cite{mylavarapu2024teleportation,roy2025teleportation}.

Our manuscript is organised as follows. In Section~\ref{sec:resource_states}, we provide a brief overview of the process of broadcasting entanglement and executing teleportation.  Therein, we also define the resource states involved in these processes. In Section~\ref{sec:gans}, we describe the structure of GAN and the evaluation metrics associated with it. In Section~\ref{sec:resource_states_cgan}, we generate resource states through GAN and compare them with the existing theoretical results. Thereafter, in Section~\ref{sec:discussion}, we provide the concluding remarks, and finally in Section~\ref{sec:outlook}, we discuss the future research efforts that can be further explored in this direction.

\section{Quantum Resource States}
\label{sec:resource_states}
Quantum resource states underpin numerous tasks in quantum information processing. In this section, we outline two pivotal protocols—broadcasting of quantum entanglement and quantum teleportation—and clarify the criteria that render a quantum state a useful resource for each. The identification and synthesis of such states is central to the GAN-based framework developed in this work.

\subsection{Resource States for Broadcasting}

Broadcasting of quantum entanglement is designed to increase the availability of entanglement across a network, effectively decomposing entanglement to generate more entangled pairs from an initial quantum resource ~\cite{buzek1997broadcasting,chatterjee2016broadcasting,mundra2019broadcasting}. Although typically demonstrated for two-qubit systems, the concept generalizes to higher-dimensional and multipartite settings, and has been extended to resources beyond entanglement, such as quantum discord and coherence ~\cite{sharma2017broadcasting,patel2024broadcasting,modi2010unified}. Quantum operations for broadcasting are usually implemented via cloning  but not always restricted to cloning \cite{jain2019asymmetric}. In this we consider broadcasting with the help of both local and nonlocal cloning.\newline

\paragraph{Local Cloning.}
Consider an entangled state $\rho_{12}$ shared between Alice and Bob. Local broadcasting introduces ancilla qubits $|\Sigma_3\rangle$, $|\Sigma_4\rangle$, followed by local cloning operations ($U_1$ at Alice's site and $U_2$ at Bob's). The output state is constructed as shown in Eq.~\eqref{eq:local_cloning_output}:
\begin{equation}
\rho'_{1234}= (U_1 \otimes U_2) (\rho_{12} \otimes |\Sigma_3\rangle \otimes |\Sigma_4\rangle).
\label{eq:local_cloning_output}
\end{equation}
Tracing out the corresponding qubits, one obtains the local marginals in Eq.~\eqref{eq:local_marginals} and nonlocal marginals in Eq.~\eqref{eq:nonlocal_marginals}:
\begin{align}
\rho'_{13} &= \mathrm{Tr}_{24}\left(\rho'_{1234}\right), &
\rho'_{24} &= \mathrm{Tr}_{13}\left(\rho'_{1234}\right), \label{eq:local_marginals} \\
\rho'_{14} &= \mathrm{Tr}_{23}\left(\rho'_{1234}\right), &
\rho'_{23} &= \mathrm{Tr}_{14}\left(\rho'_{1234}\right). \label{eq:nonlocal_marginals}
\end{align}
Entanglement is considered \textit{broadcast non-optimally} if the nonlocal pairs $\rho'_{14}, \rho'_{23}$ remain entangled~\cite{buzek1997broadcasting}. For \textit{optimal broadcasting}, it is additionally required that the local pairs $\rho'_{13}, \rho'_{24}$ are separable~\cite{chatterjee2016broadcasting,mundra2019broadcasting}.\newline

\paragraph{Nonlocal Cloning.}
When both qubits are co-located (are in the same lab), a global unitary $U_{12}$ can be applied, as shown in Eq.~\eqref{eq:nonlocal_cloning_output}:
\begin{equation}
\rho'_{1234}= U_{12} (\rho_{12} \otimes |\Sigma_3\rangle \otimes |\Sigma_4\rangle).
\label{eq:nonlocal_cloning_output}
\end{equation}
Local and nonlocal marginals are similarly defined by partial traces. The same entanglement criteria apply for optimal vs. non-optimal broadcasting.\\

\paragraph{Entanglement Detection.}
In $2 \times 2$ and $2 \times 3$ systems, the Peres-Horodecki criterion provides a necessary and sufficient test: if the partial transpose of $\rho$ yields a negative eigenvalue, the state is entangled~\cite{peres1997quantum,WernerPhysRevA.40.4277}. Here we use this criteria to detect whether the output states are entangled or not. 

\vspace{1em}
Broadcasting has also been generalized to resources beyond entanglement—such as NPT states, coherence, and Bell nonlocality—and performed using asymmetric cloning, higher-dimensional operations, and continuous-variable systems~\cite{mundra2021broadcasting,sharma2017broadcasting,adhikari2008broadcasting,jain2019asymmetric,patel2024broadcasting}. Not every mixed state is broadcastable~\cite{barnum2007generalized}; quantum mechanics fundamentally limits perfect cloning and broadcasting as compared to classical information theory~\cite{wootters1982single,barnum2007generalized}.

\subsection{Resource States for Teleportation}

Quantum teleportation exploits entanglement to transfer quantum information between parties. If Alice and Bob share a maximally entangled Bell state, such as the one in Eq.~\eqref{eq:bell_state},
\begin{equation}
|\Psi_{+}\rangle = \frac{1}{\sqrt{2}}(|00\rangle + |11\rangle),
\label{eq:bell_state}
\end{equation}
then an unknown qubit can be teleported with fidelity $F = 1$~\cite{bennett1993teleporting,horodecki1996teleportation}. More generally, the teleportation fidelity depends on the entanglement quality of the shared resource. For example, in the absence of entanglement, the maximal achievable fidelity is $F = 2/3$~\cite{horodecki1996teleportation}. Thus, any state giving $F > 2/3$ constitutes a resource for quantum teleportation.

Given a two-qubit density matrix of the general form in Eq.~\eqref{eq:general_rho},
\begin{equation}
\rho = \frac{1}{4}\left[ I_4 + \sum_{i} \sigma_i \otimes I + \sum_{j} I \otimes \sigma_j + \sum_{i,j} t_{ij} \sigma_i \otimes \sigma_j \right],
\label{eq:general_rho}
\end{equation}
(where $\sigma_i$ are Pauli matrices and $T = [t_{ij}]$ is the correlation matrix), the teleportation criterion is given by Eq.~\eqref{eq:teleportation_fidelity}:
\begin{equation}
F_\text{max}(\rho) = \frac{1}{2}\left[ 1 + \frac{1}{3} N(\rho) \right], \qquad N(\rho) = \mathrm{Tr}\left( \sqrt{T^\dagger T} \right)
\label{eq:teleportation_fidelity}
\end{equation}
States with $N(\rho) > 1$ or equivalently $F_\text{max}(\rho) > \frac{2}{3}$ provide quantum advantage~\cite{horodecki1996teleportation,chakrabarty2010teleportation,chakrabarty2011deletion,adhikari2008quantum}.Note that not every entangled state is useful for teleportation; counterexamples are documented~\cite{Horodecki1998,chakrabarty2010teleportation,chakrabarty2011deletion}. The idea of a resource state in context is also being extended for a quantum network, by quantifying networks ability to do distributed teleportation \cite{mylavarapu2023entanglement,mylavarapu2024teleportation}.

\vspace{1em}

\section{Overview of GAN}
\label{sec:gans}
A GAN consists of two neural networks in competition: a generator \(G(z)\) that maps latent noise \(z\) to candidate states, and a discriminator \(D(x)\) that distinguishes generated states from training samples \cite{Goodfellow2014,Goodfellow2020gan}. Training is framed as the minimax game shown in Eq.~\eqref{eq:gan_minimax}:
\begin{align}
\min_G \max_D V(D,G)
  &= \mathbb{E}_{x \sim P_{\mathrm{data}}}\!\bigl[\log D(x)\bigr] \nonumber\\
  &\quad+ \mathbb{E}_{z \sim P_{z}}\!\bigl[\log\!\bigl(1 - D(G(z))\bigr)\bigr].
\label{eq:gan_minimax}
\end{align}
Intuitively, training proceeds as a two-player game: the discriminator improves at distinguishing training from generated samples, while the generator adapts to produce outputs that the discriminator cannot reject. At equilibrium, the generator's output distribution matches the training distribution, enabling synthesis of new samples with the same statistical properties. For density-matrix generation, this dynamic is particularly advantageous: the discriminator implicitly learns the manifold of valid quantum states, providing a learned enforcement of physicality that complements explicit constraint terms in the loss function.

A schematic overview of our physics-informed training pipeline is shown in Figure~\ref{fig:framework_diagram}. During training, the generator maps latent samples \(z\) to candidate density matrices \(\rho_{\text{gen}} = G(z)\). The discriminator compares \(\rho_{\text{gen}}\) with training samples \(\rho_{\text{train}}\) to produce the adversarial signal, while a separate physics module computes differentiable penalties for trace-one normalization, positive semidefiniteness, and task utility (broadcasting or teleportation). The generator is updated using the resulting weighted objective (Eq.~\eqref{eq:generator_loss}). After training, the generator is used as a sampler: drawing \(z\) yields \(\rho_{\text{resource}} = G^*(z)\), a candidate resource state.

\subsection{Conditional GANs and Quantum Constraints}
\begin{figure*}[t!]
    \centering
    \includegraphics[width=\textwidth]{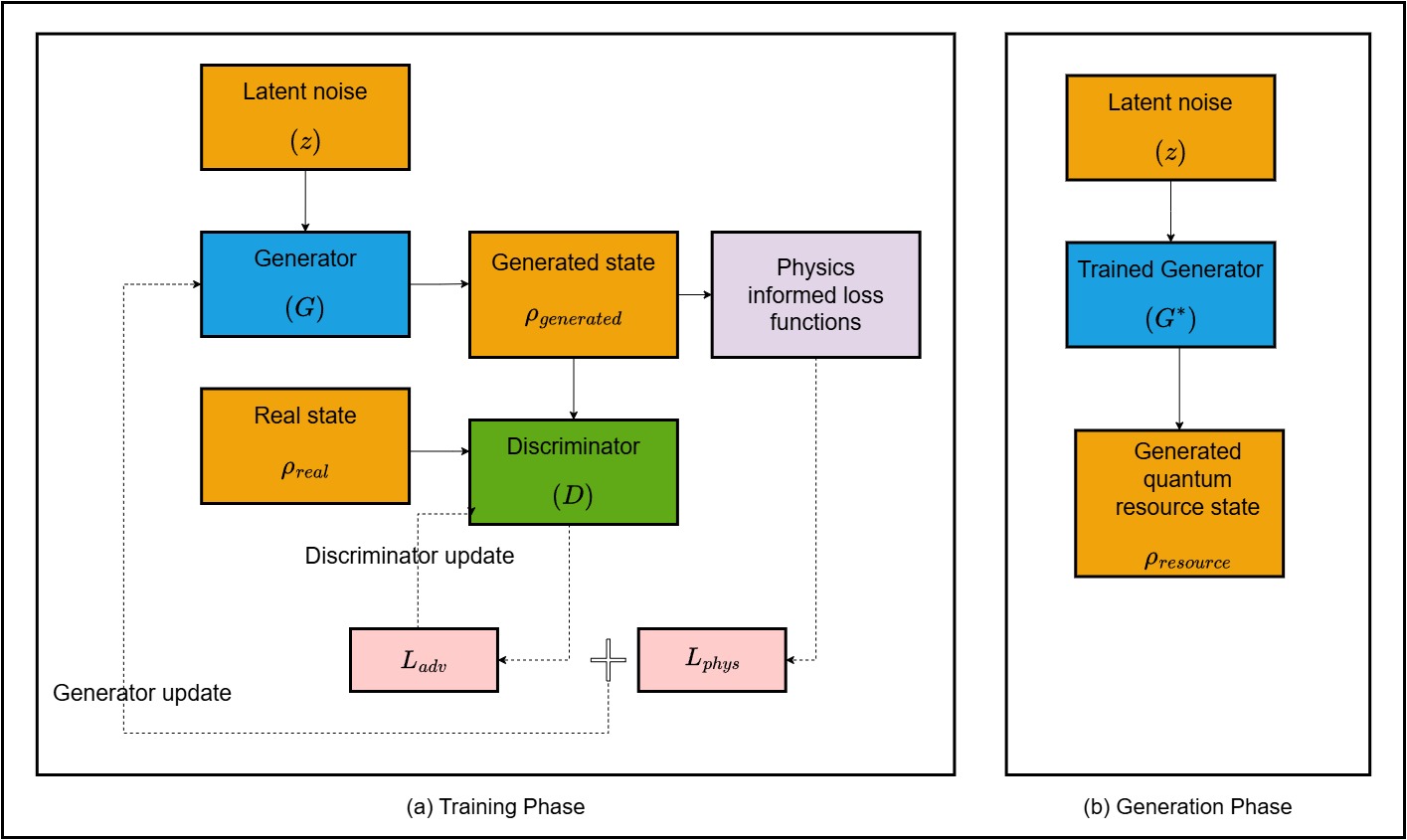}
\caption{Physics-informed CGAN framework used in this work.
(a) Training: the generator \(G\) maps latent noise \(z\) to \(\rho_{\text{gen}}\). The discriminator \(D\) compares \(\rho_{\text{gen}}\) with training samples \(\rho_{\text{train}}\), producing the adversarial term \(L_{\text{adv}}\). In parallel, a physics module computes constraint and utility terms (trace-one, PSD, and task utility), yielding the weighted objective in Eq.~\eqref{eq:generator_loss}. The discriminator and generator are updated alternately.
(b) Inference: after training, samples \(z\) are mapped by the trained generator \(G^*\) to \(\rho_{\text{resource}}\), producing candidate quantum resource states.}
    \label{fig:framework_diagram}
\end{figure*}

For quantum state generation, validity requires Hermiticity, trace-one normalization, and positive semidefiniteness. A conditional GAN (CGAN)~\cite{mirza2014conditionalgenerativeadversarialnets} extends the standard GAN by incorporating auxiliary information into both generator and discriminator, enabling targeted generation conditioned on external signals. In our setting, these signals take the form of physics-informed losses: trace loss, positive semi-definite (PSD) loss, and utility-specific losses (broadcasting vs.\ teleportation), which enforce both physical validity and task relevance. Figure~\ref{fig:framework_diagram} provides a schematic representation of the physics-informed CGAN framework used throughout this work.


\subsubsection{Generator Architectures}
We compare three generator designs that differ in how they enforce physicality. Models~1--2 impose Hermiticity and positive semidefiniteness by construction via matrix factorizations, and enforce the trace-one constraint via normalization. Model~3 outputs a Hermitian matrix directly and relies on loss penalties to learn physical validity.
\begin{itemize}[leftmargin=*, itemsep=2pt]
    \item \textit{Cholesky Decomposition} (Model 1): the network outputs the entries of a lower-triangular factor \(L\) (implemented by filling the lower triangle from a real-valued parameter vector). We assemble \(\tilde{\rho} = L L^\dagger\), which is Hermitian and PSD by construction, and enforce trace-one by \(\rho=\tilde{\rho}/\mathrm{Tr}(\tilde{\rho})\).
    \item \textit{Enhanced LDL Decomposition} (Model 2): the network outputs a unit-lower-triangular \(L\) and a diagonal \(D\) with strictly positive entries (implemented with a small \(\epsilon\) floor for numerical stability), and constructs \(\tilde{\rho}=L D L^\dagger\), followed by trace normalization.
    \item \textit{Direct Matrix Generator} (Model 3): the network outputs an unconstrained real vector that is reshaped into the real and imaginary parts of a \(4\times4\) matrix, which is then symmetrized to obtain a Hermitian \(\tilde{\rho}\). Trace-one and PSD are not guaranteed by construction and are instead encouraged purely through the explicit loss terms.
\end{itemize}
Full architectural specifications and training hyperparameters are provided in Appendix~\ref{app:reproducibility}.

\subsubsection{Adversarial Training}
The generator and discriminator are trained in competition. The discriminator is a multilayer perceptron (MLP) with Leaky Rectified Linear Unit (ReLU) activations and dropout regularization. The generator is optimized using the composite loss in Eq.~\eqref{eq:generator_loss}:
\begin{equation}
G_{\text{loss}} = L_{\text{adv}} + \lambda_1 L_{\text{trace}} + \lambda_2 L_{\text{psd}} + \lambda_3 L_{\text{task}},
\label{eq:generator_loss}
\end{equation}
where \(L_{\text{task}}\) represents teleportation fidelity loss or broadcasting separability/entanglement loss.

\subsubsection{Loss Functions and Evaluation Metrics}

\paragraph{Loss Functions:} To ensure the generator produces physically valid and task-specific quantum states, we employ a composite loss function, \(G_{\text{loss}}\). This function combines the standard adversarial loss with several penalty terms designed to enforce the required quantum properties. Key components include penalties for deviations from positive semi-definiteness (\(L_{\text{psd}}\)) and the trace-one condition (\(L_{\text{trace}}\)). Furthermore, we introduce a utility-based loss,
$L_{task}$, which for the teleportation task specifically penalizes
states that do not meet the required fidelity threshold for a
quantum advantage. The full mathematical formulation of each loss component and the final weighted-sum equation are detailed in Appendix~\ref{app:loss}.
\\

\paragraph{Evaluation Metrics:} To assess performance, we use three primary metrics to compare the set of generated states, \(\mathcal{G}\), against the set of training states, \(\mathcal{T}\).\\
\textit{b1. Accuracy:} We define accuracy as the fraction of generated states that satisfy the predefined utility criteria for the target resource class (e.g., $F_{\text{max}} > 2/3$). This metric directly evaluates the generator's success in producing useful states.\\
\textit{b2. Average Cross-Set Fidelity:} To measure state-level similarity, we compute the average fidelity between the set of generated states $\mathcal{G}$ and the set of training states $\mathcal{T}$:
\begin{equation}
F(\mathcal{G},\mathcal{T}) = \frac{1}{|\mathcal{G}||\mathcal{T}|}\sum_{\rho_g \in \mathcal{G}}\sum_{\rho_t \in \mathcal{T}} F(\rho_g, \rho_t),
\label{eq:cross_fidelity}
\end{equation}
where $F(\rho_g, \rho_t)$ is the Uhlmann fidelity between two density matrices. A score closer to the average pairwise fidelity of the training data with itself indicates a perfect match in both quality and variety. This intra-dataset fidelity serves as an ideal benchmark, and we plot it as a horizontal baseline in our results to provide a clear target for the generator's performance.\\
\textit{b3. Fréchet Inception Distance (FID):} 
To assess distributional similarity between generated and training states, we adapt the Fréchet Inception Distance~\cite{heusel2018ganstrainedtimescaleupdate}.
Feature embeddings $\phi(\rho)$ are computed as Pauli-expectation vectors: for a two-qubit state, $\phi(\rho) = (\langle \sigma_i \otimes \sigma_j \rangle)_{i,j=0}^{3}$, yielding a 16-dimensional real vector. The training and generated feature distributions are modeled as multivariate Gaussians, with means 
$(\mu_{\mathcal{G}}, \mu_{\mathcal{T}})$ and covariances $(\Sigma_{\mathcal{G}}, \Sigma_{\mathcal{T}})$. 
The FID score is then given by Eq.~\eqref{eq:fid_score}:
    \begin{equation}
    \begin{split}
    \text{FID}(\mathcal{G}, \mathcal{T}) ={}& \|\mu_{\mathcal{G}} - \mu_{\mathcal{T}}\|^2 \\
    & + \mathrm{Tr}\!\left(\Sigma_{\mathcal{G}} + \Sigma_{\mathcal{T}}
    - 2\sqrt{\Sigma_{\mathcal{G}}\Sigma_{\mathcal{T}}}\right).
    \end{split}
    \label{eq:fid_score}
    \end{equation}
Lower values indicate that the generator not only produces valid states but also 
captures the diversity and covariance structure of the target distribution. 
Unlike average fidelity, which measures pairwise similarity at the state level, FID is sensitive to coverage and mode collapse, providing complementary insight into how well the GAN reproduces the overall training distribution.

\section{Generation of Resource States through CGAN}
\label{sec:resource_states_cgan}
Having established the CGAN framework and evaluation methodology, we now apply it to generate quantum resource states for broadcasting and teleportation. In this section, we discuss the generation of two different types of resource states (broadcastable and teleportable) through the three generator architectures described above. We do these for two different types of states, namely Bell diagonal and Werner like states. In the later part of the section we also compare the states generated  with the known theoretical ranges for these classes of states. Before we go to the first subsection where we analyze and compare between the generation process, we give a brief description of the inputs that we have taken.\par

\paragraph{Werner-like states.} These states are defined by the Bloch vector and correlation matrix given in Eq.~\eqref{eq:werner_like_def}:
\begin{align}
    \rho_{12}^w &= \{ \vec{x}^w, \vec{x}^w, T^w \}, \nonumber \\
    \vec{x}^w &= \{ 0, 0, p (\alpha^2 - \beta^2) \}, \nonumber \\
    T^w &= \text{diag}(2p\alpha\beta, -2p\alpha\beta, p),
    \label{eq:werner_like_def}
\end{align}
    with the conditions \( \alpha^2 + \beta^2 = 1 \) and \( 0 \leq p \leq 1 \). Note that the diagonal matrix \( M = \text{diag}(\dots) \) has diagonal elements specified inside the brackets.
\paragraph{Bell-diagonal states.} These states can be expressed via Eq.~\eqref{eq:bell_diag_def}, where the Bloch vector is null:
\begin{align}
    \rho_{12}^b &= \{ \vec{0}, \vec{0}, T^b \}, \nonumber \\
    T^b &= \text{diag}(c_1, c_2, c_3),
    \label{eq:bell_diag_def}
\end{align}
with \( -1 \leq c_i \leq 1 \). The eigenvalues of \( \rho_{12}^b \) are given by Eq.~\eqref{eq:bell_diag_eigenvalues}:
\begin{equation}
\lambda_{mn} = \frac{1}{4} \left[ 1 + (-1)^m c_1 - (-1)^{m+n} c_2 + (-1)^n c_3 \right],
\label{eq:bell_diag_eigenvalues}
\end{equation}

\subsection{Generation and Analysis}
In the following analysis, Fig.~\ref{fig:accuracy_comp_g1g2} reports task accuracy, Fig.~\ref{fig:fidelity_comp_g1g2} reports average cross-set fidelity, and Fig.~\ref{fig:fid_comp_g1g2} reports Fréchet Inception Distance (FID). Each figure presents (1)~Bell-diagonal and (2)~Werner-like states, with three subplots per row: (a)~local broadcasting, (b)~non-local broadcasting, and (c)~teleportation. Complete numerical results are tabulated in Appendix~\ref{app:results}.

\paragraph{Model snapshot and epoch selection.}
All scatter plots showing generated states in the main text (Figs.~\ref{fig:theoretical_bell_diagonal_regions} and~\ref{fig:theoretical_werner_like_regions}) use the same model snapshot: Model 1 (Cholesky Decomposition) at epoch 3000, trained on dataset size 2000. This epoch was selected to enable consistent comparison across all six task-state combinations, representing a point where Cholesky achieves near-optimal performance across most metrics while avoiding the degraded accuracy and elevated FID observed at earlier epochs (see Appendix~\ref{app:counter_examples}). We emphasize that GAN training does not converge monotonically~\cite{Goodfellow2014,salimans2016improved}; significant epoch-to-epoch performance variation is visible in several settings (Figs.~\ref{fig:accuracy_comp_g1g2}, \ref{fig:fidelity_comp_g1g2}, and~\ref{fig:fid_comp_g1g2}), making a single common evaluation epoch preferable to selecting different epochs per task.\\

Across all experiments, a clear distinction emerged between constraint-enforcing and unconstrained generator architectures. Decomposition-based models (Cholesky, Enhanced LDL) consistently achieved rapid convergence and stable accuracy, with average cross-set fidelity scores closely tracking theoretical benchmarks for both Werner-like and Bell-diagonal states. These models, by construction, maintained physical validity throughout training, reducing the need for heavy penalties for auxiliary losses. However, their sensitivity to hyperparameter tuning also introduced risks of mode collapse~\cite{Durall2020CombatingMC} in smaller training sets.

In contrast, the direct generation model demonstrated greater robustness to hyperparameter variation but consistently lagged behind decomposition-based approaches in distribution-level measures such as Fréchet Inception Distance (FID) across all studied tasks. This indicates that while unconstrained models offer flexibility and adaptability, they did not surpass decomposition models in FID within the two-qubit scenarios considered here.

This performance gap is a key finding. It demonstrates that for the well-structured two-qubit problems studied here, embedding physical constraints directly into the generator's architecture is a far more effective and stable strategy than relying on penalty-based learning. While the flexibility of the direct model remains appealing for future, less-understood problems, our results underscore the significant value of architecturally-enforced constraints when the underlying physics is known.

Comparison with theoretical boundaries further validates the generative process. Scatter plots of generated Bell-diagonal states (Fig.~\ref{fig:theoretical_bell_diagonal_regions}) confirmed confinement to the appropriate regions, with the brown wireframe depicting the octahedron of inseparable states and the teal shaded areas marking the resource subregions at each vertex for broadcastable and teleportable states. Similarly, the Werner-like states generated (Fig.~\ref{fig:theoretical_werner_like_regions}) adhered to the theoretical boundaries, with the overwhelming majority of useful states (blue dots) falling within the green entangled region. The small number of orange dots appearing in the green region represent minor classification inaccuracies, while the overall distribution strongly reproduces the known analytical boundary $p(1+4\alpha\beta) > 1$ for teleportability (see Appendix~\ref{app:teleportation_ppt_equivalence}). These findings establish that the proposed CGAN framework does not merely memorize training samples but generalizes to produce physically valid states consistent with analytic criteria.

Figure~\ref{fig:accuracy_comp_g1g2} summarizes the task-accuracy trajectories during training.
Across both state families, the constraint-enforcing generators (Cholesky, Enhanced LDL) reach high accuracy early and remain comparatively stable across epochs, indicating that once the generator learns to stay within the physically valid manifold, satisfying the task utility condition becomes reliably achievable. In contrast, the Direct Matrix generator exhibits substantially lower accuracy across tasks and a noticeably slower rise,
consistent with the additional burden of learning physical validity and task structure purely through penalty terms.

A second qualitative trend visible in Fig.~\ref{fig:accuracy_comp_g1g2} is task dependence:
broadcasting tasks exhibit greater variability across epochs than teleportation, which is expected because the broadcasting labels are defined via separability/entanglement constraints on cloned output marginals rather than a single scalar threshold. For both Bell-diagonal and Werner-like settings, the decomposition-based models maintain accuracy curves that are consistently closer to the desired accuracy than the direct model, supporting the conclusion that architectural
constraint enforcement improves stability for these two-qubit resource-generation tasks.
\begin{figure*}[!ht]
    \centering
    \subfloat[Bell-diagonal states\label{fig:acc_bell}]{
        \includegraphics[width=0.99\textwidth]{Combined_Bell_Diagonal_Accuracy_Plots.png}}
    \hfill
    \subfloat[Werner-like states\label{fig:acc_werner}]{
        \includegraphics[width=0.99\textwidth]{Combined_Werner-like_Accuracy_Plots.png}}

  \caption{Accuracy comparison across generator architectures (training size = 500).
  (1)~Bell-diagonal states; (2)~Werner-like states. Subplots: (a)~local broadcasting, (b)~non-local broadcasting, (c)~teleportation.}
  \label{fig:accuracy_comp_g1g2}
\end{figure*}

Figure~\ref{fig:fidelity_comp_g1g2} compares the models on average cross-set fidelity, which measures how well the distribution of generated states matches the training data. For both broadcastable and teleportable Werner-like states, the decomposition-based Cholesky and Enhanced LDL models perform exceptionally well, achieving and maintaining high fidelity. The Direct Matrix generator, in contrast, shows a significant drop in fidelity, indicating a failure to correctly learn the target distribution for these more complex states.

In Fig.~\ref{fig:fidelity_comp_g1g2}, the dashed horizontal baseline denotes the average fidelity of the training set with itself, which serves as an internal reference for the typical pairwise similarity scale within the target distribution. Fidelity curves that approach this baseline indicate that the generator is producing states that are statistically similar to the
training data at the state level, while sustained gaps indicate a mismatch in either typical state quality or coverage.

Qualitatively, the decomposition-based models track the training baseline more closely across both Bell-diagonal and Werner-like tasks, with particularly strong agreement in the Werner-like broadcasting cases where the curves remain high and stable. By comparison, the Direct Matrix Generator shows larger deviations in several tasks, consistent with reduced agreement at the state level. These observations align with the accuracy trends: models that enforce positivity and trace structurally not only satisfy the utility criteria more often, but also tend to generate states that remain closer to the training distribution under the fidelity metric.

\begin{figure*}[!ht]
    \centering
    \subfloat[Bell-diagonal states\label{fig:fid_bell}]{
        \includegraphics[width=0.99\textwidth]{Combined_Bell_Diagonal_Fidelity.png}}
    \hfill
    \subfloat[Werner-like states\label{fig:fid_werner}]{
        \includegraphics[width=0.99\textwidth]{Combined_Wernerlike_Fidelity.png}}

\caption{Average cross-set fidelity between generated and training states (training size $=1000$). (1)~Bell-diagonal states; (2)~Werner-like states. Subplots: (a)~local broadcasting, (b)~non-local broadcasting, (c)~teleportation. The dashed red line indicates the ideal benchmark—the average fidelity of the training data with itself.}
\label{fig:fidelity_comp_g1g2}
\end{figure*}

Finally, Figure~\ref{fig:fid_comp_g1g2} presents Fréchet Inception Distance (FID) between generated and training distributions. Decomposition-based models consistently achieve lower final FID scores, indicating closer alignment with training sets. The Direct Matrix Generator sometimes exhibits visually stable trajectories (notably in nonlocal Bell-diagonal and teleportable Werner-like tasks), but its reported endpoint FID values remain higher, confirming the quantitative advantage of decomposition models. The Direct Matrix Generator also exhibits FID instability in specific cases: for Werner-like local broadcasting, FID diverges mid-training and fails to recover, while for Bell-diagonal local broadcasting, FID increases sharply towards the end of training. This instability likely stems from violations of positive semidefiniteness that distort the feature covariance.

FID provides a complementary, distribution-level view: lower values indicate that the generated set reproduces not only typical samples but also the mean and covariance structure of the feature representation used (Pauli-expectation embeddings).
Thus, FID is particularly sensitive to reduced diversity or partial mode collapse, even in cases where pointwise fidelity can remain moderate.

The trajectories in Fig.~\ref{fig:fid_comp_g1g2} show that decomposition-based generators typically converge to lower final FID values than the Direct Matrix Generator across tasks, indicating closer alignment with the target distribution at the level of first- and second-order statistics. In several settings the direct model exhibits comparatively flatter or slowly improving curves, but with higher endpoints, consistent with a residual distribution mismatch. Taken together with Fig.~\ref{fig:fidelity_comp_g1g2}, these trends support the interpretation that enforcing physical structure in the generator improves both task success and distributional coverage for the two-qubit cases studied here.

\begin{figure*}[!ht]
    \centering
   \subfloat[Bell-diagonal states\label{fig:frechet_bell}]{
        \includegraphics[width=0.99\textwidth]{Combined_Bell_Diagonal_FID.png}}
    \hfill
    \subfloat[Werner-like states\label{fig:frechet_werner}]{
        \includegraphics[width=0.99\textwidth]{Combined_Werner_Like_FID.png}}

  \caption{Fréchet Inception Distance (FID) between generated and training state distributions (training size = 2000).
  (1)~Bell-diagonal states; (2)~Werner-like states. Subplots: (a)~local broadcasting, (b)~non-local broadcasting, (c)~teleportation. Lower FID indicates closer distributional match.}
  \label{fig:fid_comp_g1g2}
\end{figure*}

\subsection{Comparison with Theoretical Results}

\begin{figure*}
  \centering
  \subfloat[\label{fig:sub_theo_bell_local}]{
    \includegraphics[width=0.32\textwidth]{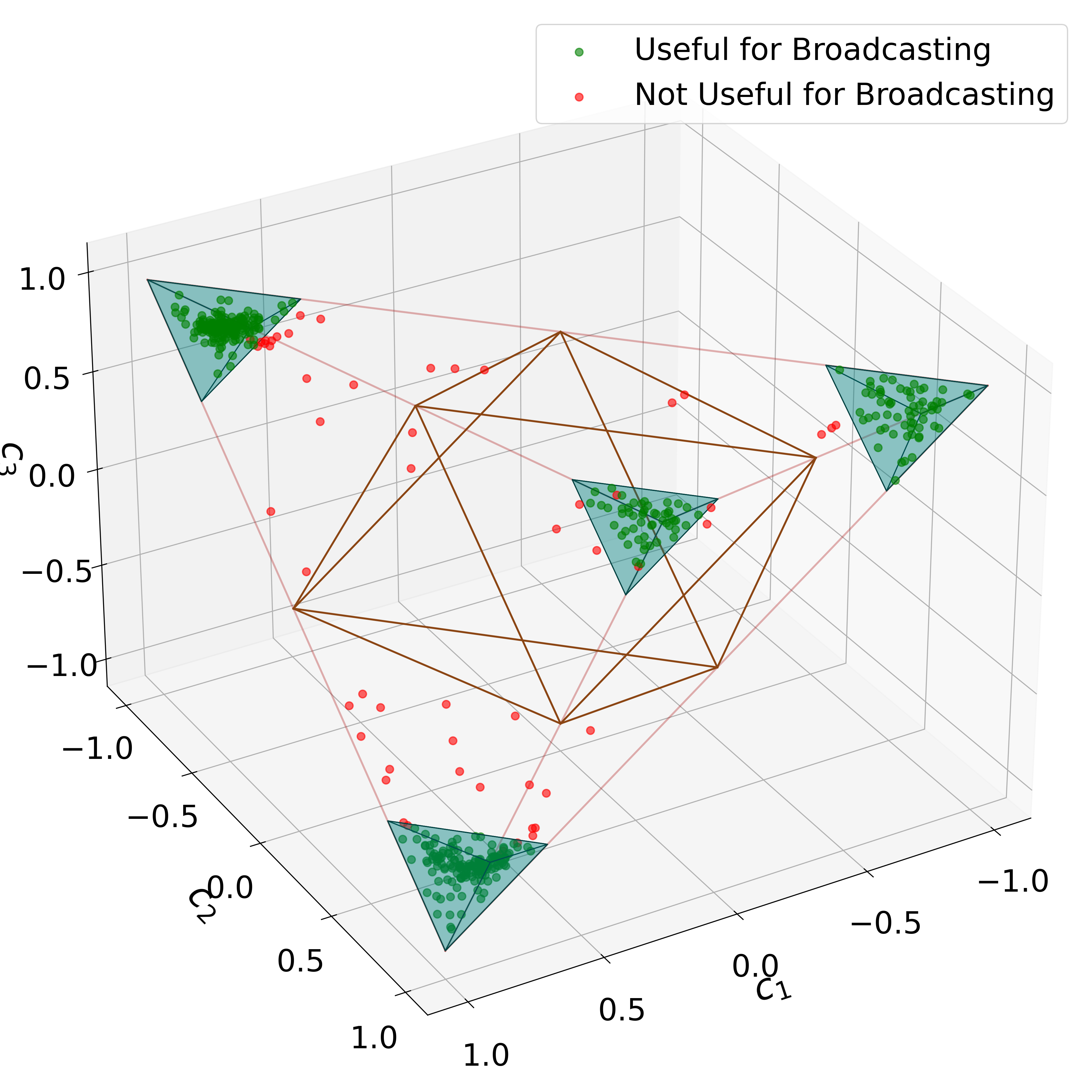}}
  \hfill
  \subfloat[\label{fig:sub_theo_bell_nonlocal}]{
    \includegraphics[width=0.32\textwidth]{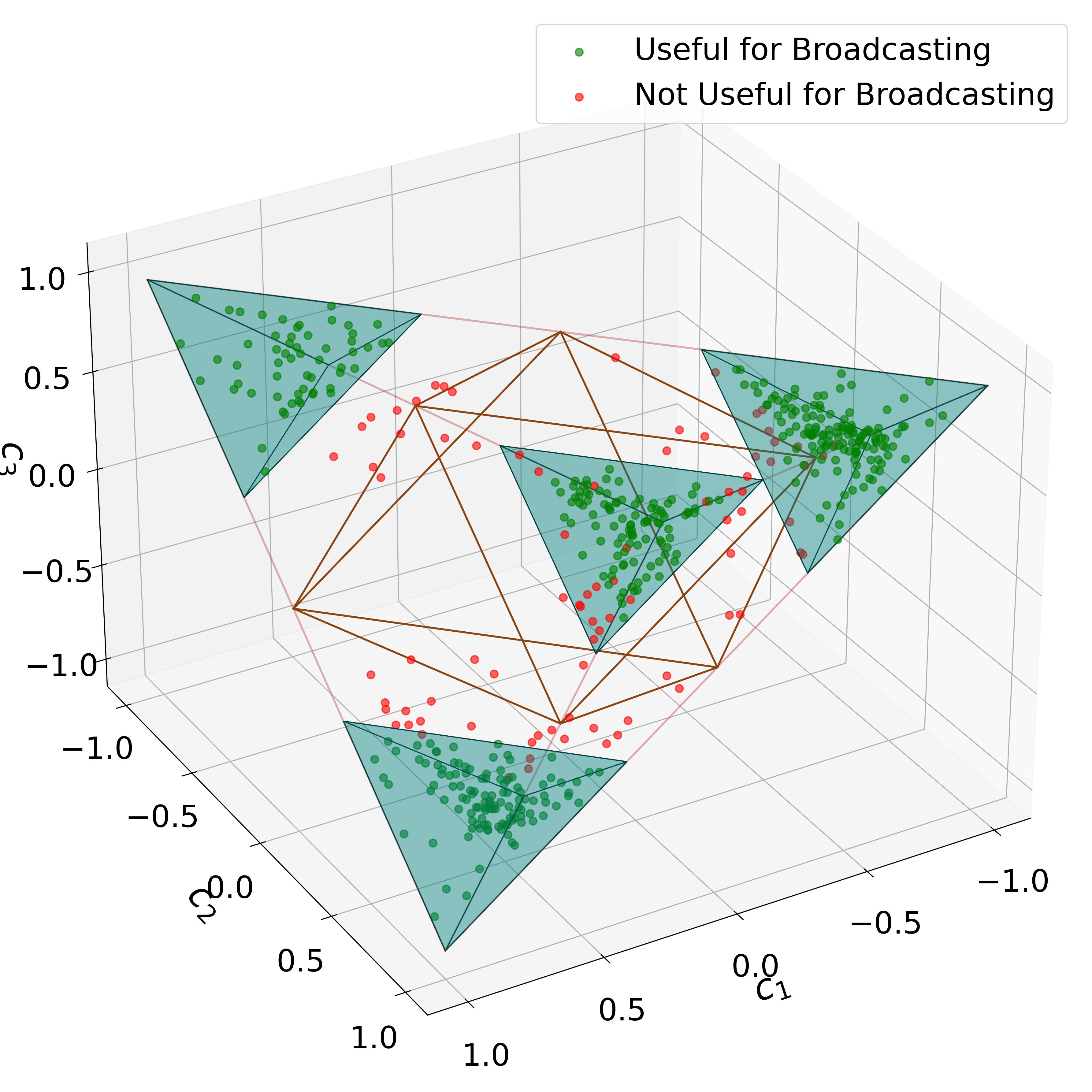}}
  \hfill
  \subfloat[\label{fig:sub_theo_bell_tele}]{
    \includegraphics[width=0.32\textwidth]{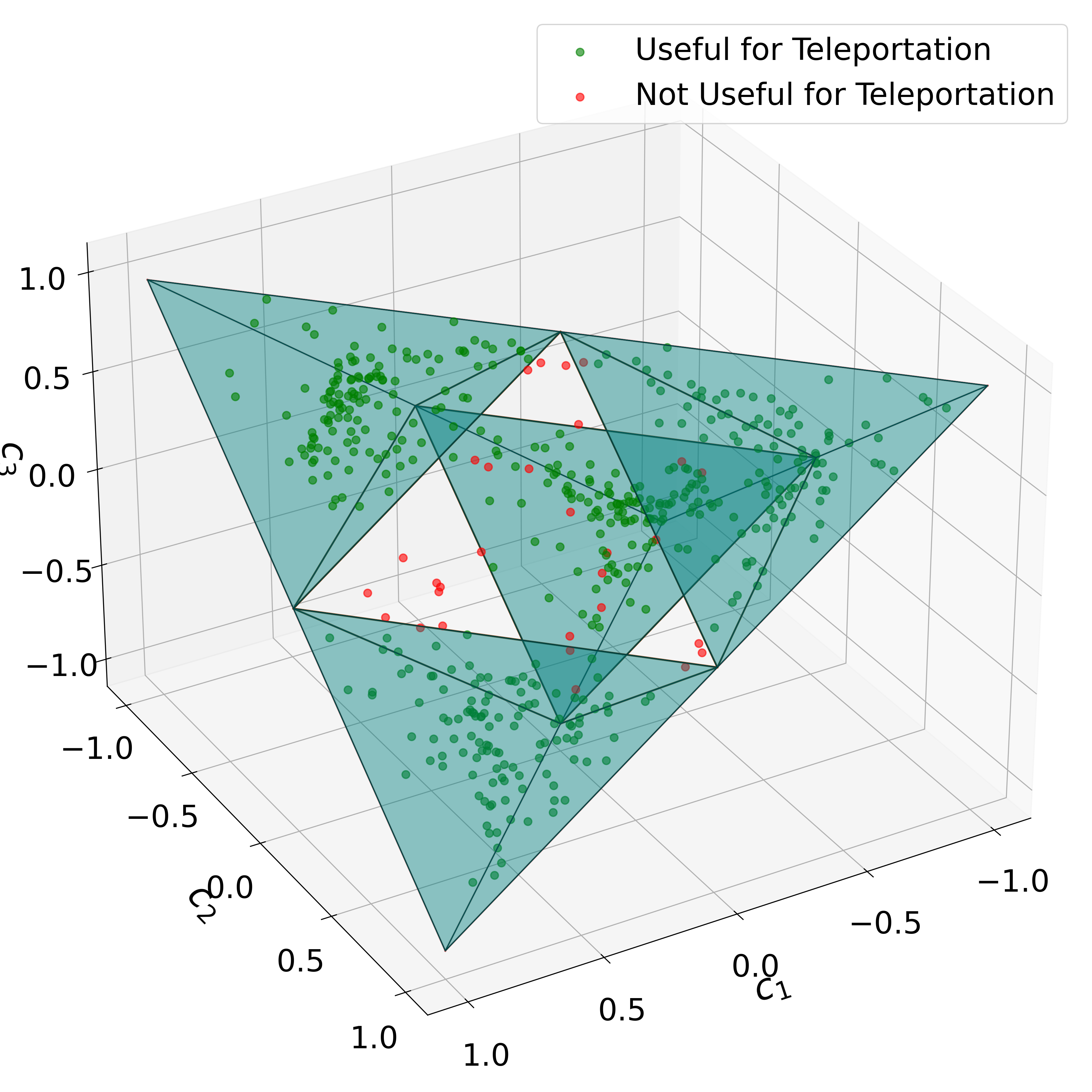}}

  \caption{Generated Bell-diagonal states plotted in $(c_1, c_2, c_3)$ parameter space.
  Subplots: (1)~local broadcasting, (2)~non-local broadcasting, (3)~teleportation.
  Brown wireframe: octahedron of inseparable states within the tetrahedron of all valid Bell-diagonal states. Teal regions: theoretical parameter ranges where states are resources for each task. Dots show generator outputs (candidates the model produced as potentially useful) labeled by post-hoc verification: green/red indicate states that, upon verification, meet/fail the resource criterion.
  All states generated by Model 1 (Cholesky Decomposition) at epoch 3000, trained with $N=2000$.}
  \label{fig:theoretical_bell_diagonal_regions}
\end{figure*}

Figure~\ref{fig:theoretical_bell_diagonal_regions} shows Bell-diagonal states generated and plotted using the trained models. For Bell-diagonal states, when we plot $c_1, c_2, c_3$, all valid states lie within a tetrahedron with vertices: A : $(1,-1,1)$, B : $(-1,1,1)$, C : $(1,1,-1)$, D : $(-1,-1,-1)$. In these plots, the brown wireframe depicts the boundary of the octahedron of inseparable (entangled) Bell-diagonal states, which lies inside the larger tetrahedron of all valid Bell-diagonal states. The teal shaded subregions near each vertex indicate the theoretical parameter ranges where states serve as resources for the respective task. Generated states classified as useful are shown in green, while those not meeting the resource criterion are shown in red.

All three subplots of Figure~\ref{fig:theoretical_bell_diagonal_regions} show states generated using Model 2 (Enhanced LDL) with training size 2000.

\textit{(1) Local broadcasting:} States useful for broadcasting of quantum entanglement using local cloning, concentrated around the tetrahedron vertices. The teal region corresponds to the following vertex coordinates:
\begin{alignat*}{2}
  &\text{D:}\;(-1,-1,-1),\; &&(-\tfrac{5}{8},-\tfrac{5}{8},-1),\;(-1,-\tfrac{5}{8},-\tfrac{5}{8}),\;(-\tfrac{5}{8},-1,-\tfrac{5}{8})\\
  &\text{C:}\;(1,1,-1),\; &&(\tfrac{5}{8},\tfrac{5}{8},-1),\;(1,\tfrac{5}{8},-\tfrac{5}{8}),\;(\tfrac{5}{8},1,-\tfrac{5}{8})\\
  &\text{B:}\;(-1,1,1),\; &&(-\tfrac{5}{8},\tfrac{5}{8},1),\;(-1,\tfrac{5}{8},\tfrac{5}{8}),\;(-\tfrac{5}{8},1,\tfrac{5}{8})\\
  &\text{A:}\;(1,-1,1),\; &&(\tfrac{5}{8},-\tfrac{5}{8},1),\;(1,-\tfrac{5}{8},\tfrac{5}{8}),\;(\tfrac{5}{8},-1,\tfrac{5}{8})
\end{alignat*}

\textit{(2) Non-local broadcasting:} States useful for broadcasting of quantum entanglement using non-local cloning, spanning a larger area around the vertices compared to local broadcasting. The teal region corresponds to:
\begin{alignat*}{2}
  &\text{D:}\;(-1,-1,-1),\; &&(-\tfrac{1}{3},-\tfrac{1}{3},-1),\;(-1,-\tfrac{1}{3},-\tfrac{1}{3}),\;(-\tfrac{1}{3},-1,-\tfrac{1}{3})\\
  &\text{C:}\;(1,1,-1),\; &&(\tfrac{1}{3},\tfrac{1}{3},-1),\;(1,\tfrac{1}{3},-\tfrac{1}{3}),\;(\tfrac{1}{3},1,-\tfrac{1}{3})\\
  &\text{B:}\;(-1,1,1),\; &&(-\tfrac{1}{3},\tfrac{1}{3},1),\;(-1,\tfrac{1}{3},\tfrac{1}{3}),\;(-\tfrac{1}{3},1,\tfrac{1}{3})\\
  &\text{A:}\;(1,-1,1),\; &&(\tfrac{1}{3},-\tfrac{1}{3},1),\;(1,-\tfrac{1}{3},\tfrac{1}{3}),\;(\tfrac{1}{3},-1,\tfrac{1}{3})
\end{alignat*}

\textit{(3) Teleportation:} States useful for teleportation, spanning specific regions within the tetrahedron:
\begin{alignat*}{2}
  &\text{D:}\;(-1,-1,-1),\; &&(-1,0,0),\;(0,-1,0),\;(0,0,-1)\\
  &\text{C:}\;(1,1,-1),\; &&(0,1,0),\;(1,0,0),\;(0,0,-1)\\
  &\text{B:}\;(-1,1,1),\; &&(-1,0,0),\;(0,1,0),\;(0,0,1)\\
  &\text{A:}\;(1,-1,1),\; &&(0,-1,0),\;(0,0,1),\;(1,0,0)
\end{alignat*}

\begin{figure*}
  \centering
  \subfloat[\label{fig:sub_theo_werner_local}]{
    \includegraphics[width=0.32\textwidth]{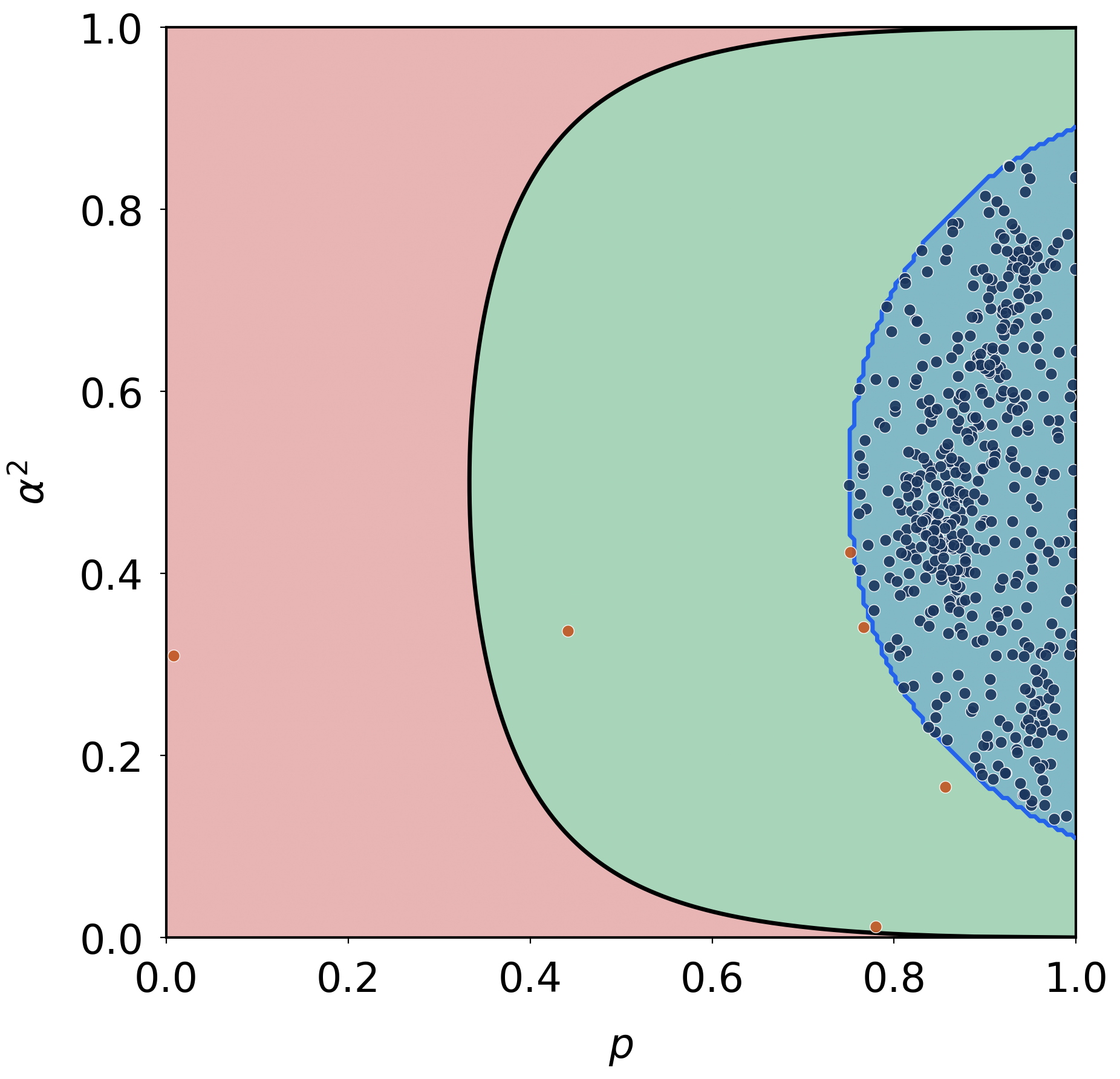}}
  \hfill
  \subfloat[\label{fig:sub_theo_werner_nonlocal}]{
    \includegraphics[width=0.32\textwidth]{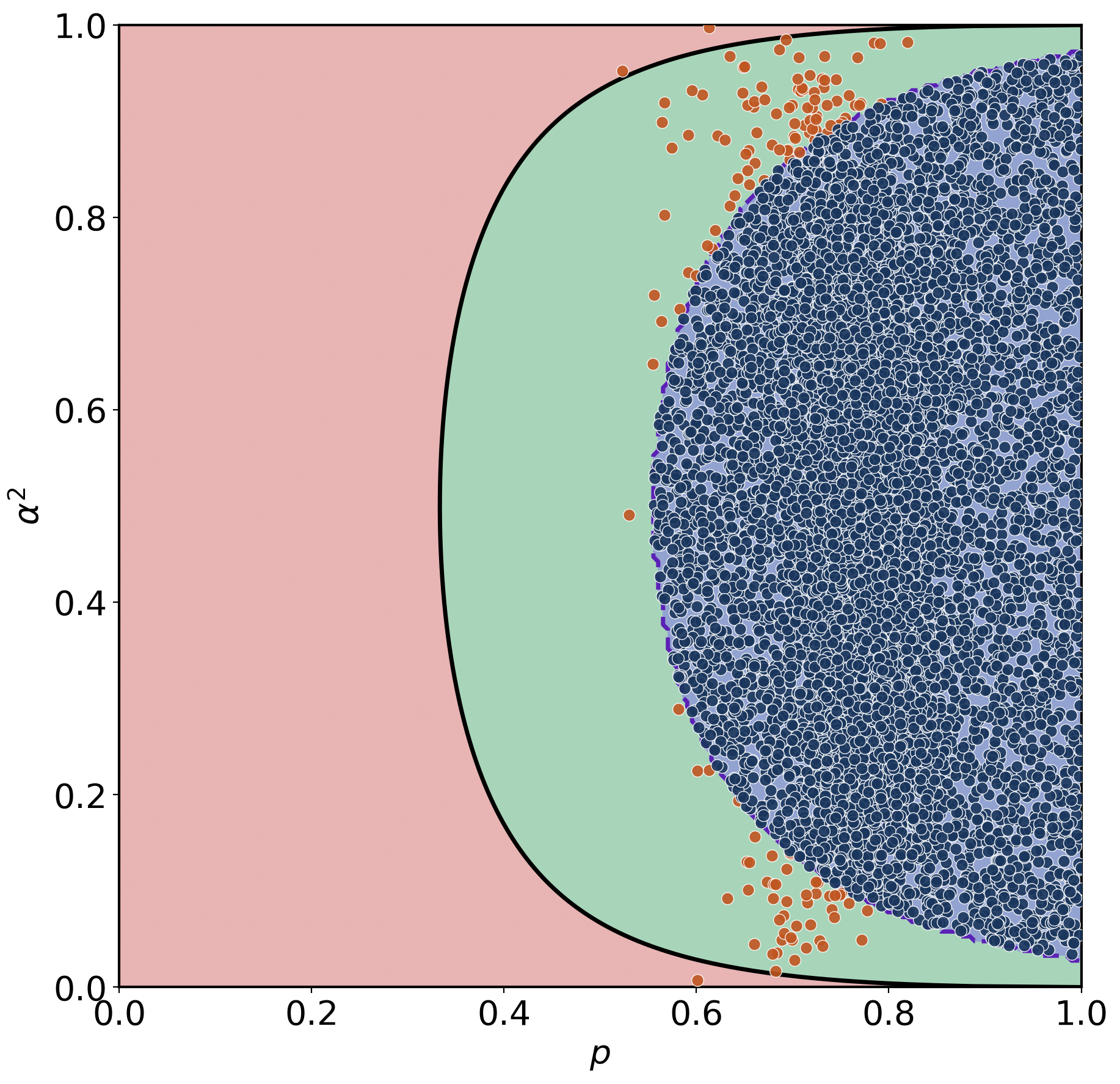}}
  \hfill
  \subfloat[\label{fig:sub_theo_werner_tele}]{
    \includegraphics[width=0.32\textwidth]{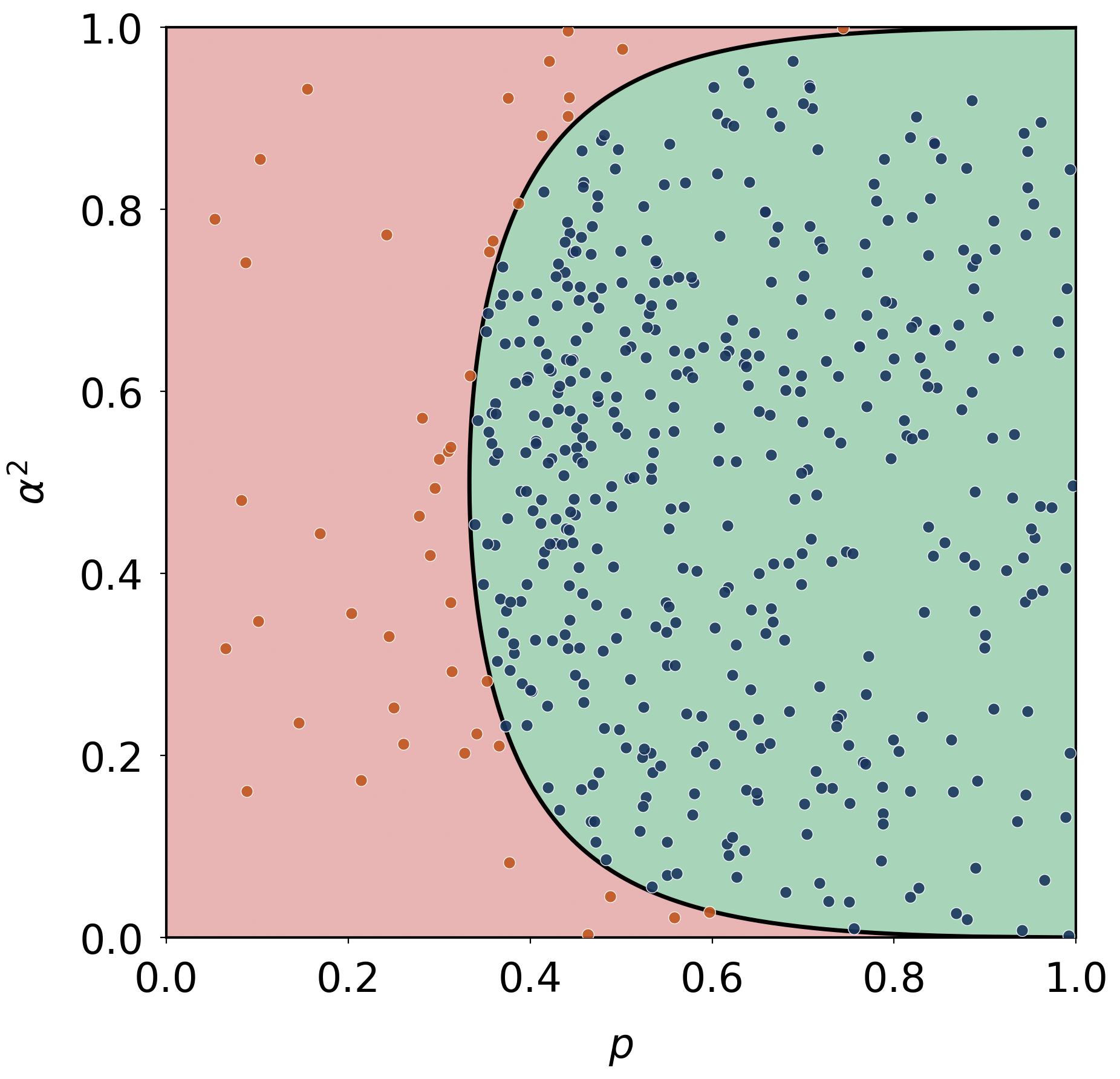}}

  \caption{Generated Werner-like states plotted in $(p, \alpha)$ parameter space.
  Subplots: (1)~local broadcasting, (2)~non-local broadcasting, (3)~teleportation.
  Green background: inseparable (entangled) states; pink background: separable region. For (1--2), green indicates broadcastable states; for (3), states useful for teleportation ($p(1+4\alpha\beta) > 1$). Dots show generator outputs (candidates the model produced as potentially useful) labeled by post-hoc verification: blue/orange indicate states that, upon verification, meet/fail the resource criterion.
  All states generated by Model 1 (Cholesky Decomposition) at epoch 3000, trained with $N=2000$.}
  \label{fig:theoretical_werner_like_regions}
\end{figure*}

Figure~\ref{fig:theoretical_werner_like_regions} presents Werner-like states generated and plotted using the trained models. In these plots, the pink background indicates the region of separable states, while the green background marks the region of inseparable (entangled) states. For the broadcasting tasks (1) and (2), the green region represents all entangled states from which broadcasting is possible; for teleportation (3), the green region coincides with states useful for teleportation. Generated states that meet the resource criterion are shown as blue dots, while those that do not are shown as orange dots; states lying in the ``wrong'' background region thus represent classification errors by the model.

All three subplots show states generated using Model 1 (Cholesky) with training size 2000: (1)~states useful for broadcasting of quantum entanglement using local cloning, (2)~states useful for broadcasting of quantum entanglement using non-local cloning, and (3)~states useful for teleportation, satisfying the utility criterion $p(1+4\alpha\beta) > 1$.

\section{Discussion and Conclusion}
\label{sec:discussion}

The results presented establish a systematic method for generating quantum states tailored to specific tasks. In this work we have chosen these tasks to be broadcasting of quantum entanglement and teleportation. The central finding is the clear performance gap between architecturally constrained generators (Cholesky, LDL) and an unconstrained, direct-output model. For the well-defined two-qubit systems studied, embedding physical constraints directly into the generator's design proved to be a more effective and stable learning strategy than relying on penalty-based loss functions to enforce validity. This suggests that for problems where the underlying physical structure is known, leveraging that structure within the model architecture is paramount.

Furthermore, our framework's ability to reproduce the known theoretical boundaries for teleportable and broadcastable states (Figs.~\ref{fig:theoretical_bell_diagonal_regions} and~\ref{fig:theoretical_werner_like_regions}) serves as a critical validation. The visual alignment of generated states with theoretical subregions---bounded by the brown wireframe octahedron of inseparable states for Bell-diagonal states and the entanglement boundary for Werner-like states---confirms that the CGAN does not merely memorize the training data but learns a generalized representation of the target resource state manifold. The direct generator, while less accurate, demonstrated robustness to hyperparameter variations, indicating its potential utility in more exploratory scenarios where physical constraints are less understood or analytically intractable, and where flexibility is more valuable than guaranteed precision. Our results is about generating task specific quantum resources in particular to the context of quantum communication which can act as the backbone of generation of resourcefull network \cite{mylavarapu2024teleportation,roy2025teleportation}. 

The present study is restricted to two-qubit systems and two specific state families (Werner-like and Bell-diagonal), for which analytical criteria exist. Whether the performance advantage of decomposition-based generators persists for higher-dimensional or less-structured state spaces remains an open question. Additionally, the framework assumes access to a well-defined utility function; tasks lacking computable metrics would require alternative formulations. These observations motivate several directions for extending and refining the framework.

\section{Future Outlook}
\label{sec:outlook}

Our results establish that physics-informed adversarial learning can generate families of valid quantum resource states tuned to protocol-specific utilities. The immediate challenge is to extend this approach beyond two-qubit systems. While computational scaling is favorable ($O(d^3)$ for validation), defining computable utilities for multipartite entanglement~\cite{horodecki2009quantum}, quantum metrology~\cite{giovannetti2011advances}, and error correction~\cite{terhal2015quantum} remains non-trivial. A natural extension is to replace the task-specific loss with other computable metrics—for instance, Quantum Fisher Information for metrology or the Geometric Measure of Entanglement for multipartite state discovery—enabling generative exploration where closed-form criteria are unknown.

A complementary direction is architectural: hybrid generators that couple decomposition-based structure with learnable residual components may balance physical faithfulness and expressivity. Embedding known symmetries and conservation laws~\cite{cohen2019gauge} could further reduce sample complexity. Finally, interfacing the framework with quantum-network simulators such as NetSquid~\cite{coopmans2021netsquid} would permit testing under realistic noise and topology models, providing a bridge between numerical design and near-term experiments.

\noindent \textit{Acknowledgment:} I.C. acknowledges the support  from the Ministry of Electronics and Information Technology (MeitY), Government of India, under Grant No. 4(3)/2024-ITEA. I.C. thanks Soumit Roy for useful discussions.

\noindent \textit{Data Availability :} \href{https://github.com/2Shabby/broadcastability_ml_project}{GitHub Repository for Project}
\bibliographystyle{apsrev4-2}

\bibliography{references}

@book{peres1997quantum,
  title     = {Quantum theory: concepts and methods},
  author    = {Peres, Asher},
  volume    = {72},
  year      = {1997},
  publisher = {Springer},
  doi       = {10.1007/0-306-47120-5},
}

@article{goodfellow2014generative,
  title={Generative adversarial nets},
  author={Goodfellow, Ian J and Pouget-Abadie, Jean and Mirza, Mehdi and Xu, Bing and Warde-Farley, David and Ozair, Sherjil and Courville, Aaron and Bengio, Yoshua},
  journal={Advances in neural information processing systems},
  volume={27},
  year={2014}
}

@inproceedings{salimans2016improved,
  title={Improved techniques for training {GAN}s},
  author={Salimans, Tim and Goodfellow, Ian and Zaremba, Wojciech and Cheung, Vicki and Radford, Alec and Chen, Xi},
  booktitle={Advances in Neural Information Processing Systems},
  volume={29},
  pages={2234--2242},
  year={2016}
}

@book{nielsen2010quantum,
  title={Quantum computation and quantum information},
  author={Nielsen, Michael A and Chuang, Isaac L},
  year={2010},
  publisher={Cambridge university press},
  doi = {10.1017/CBO9780511976667}
}

@article{rem2019identifying,
  title={Identifying quantum phase transitions using artificial neural networks on experimental data},
  author={Rem, Benno S and K{\"a}ming, Niklas and Tarnowski, Matthias and Asteria, Luca and Fl{\"a}schner, Nick and Becker, Christoph and Sengstock, Klaus and Weitenberg, Christof},
  journal={Nat. Phys.},
  volume={15},
  number={9},
  pages={917--920},
  year={2019},
  publisher={Nature Publishing Group UK London},
  doi = {10.1038/s41567-019-0554-0}
}

@article{sentis2015quantum,
  title={Quantum learning of coherent states},
  author={Sent{\'\i}s, Gael and Gu{\c{t}}{\u{a}}, M{\u{a}}d{\u{a}}lin and Adesso, Gerardo},
  journal={EPJ Quant. Tech.},
  volume={2},
  number={1},
  pages={1--22},
  year={2015},
  publisher={Springer}, 
  doi = {10.1140/epjqt/s40507-015-0030-4}
}

@article{PhysRevResearch.2.023150,
  title = {Neural-network approach for identifying nonclassicality from click-counting data},
  author = {Gebhart, Valentin and Bohmann, Martin},
  journal = {Phys. Rev. Res.},
  volume = {2},
  issue = {2},
  pages = {023150},
  numpages = {9},
  year = {2020},
  month = {May},
  publisher = {American Physical Society},
  doi = {10.1103/PhysRevResearch.2.023150},
  url = {https://link.aps.org/doi/10.1103/PhysRevResearch.2.023150}
}

@article{ma2018transforming,
  title={Transforming Bell’s inequalities into state classifiers with machine learning},
  author={Ma, Yue-Chi and Yung, Man-Hong},
  journal={npj Quant. Inf.},
  volume={4},
  number={1},
  pages={34},
  year={2018},
  publisher={Nature Publishing Group UK London},
  doi = {10.1038/s41534-018-0081-3}
}

@article{harney2020entanglement,
  title={Entanglement classification via neural network quantum states},
  author={Harney, Cillian and Pirandola, Stefano and Ferraro, Alessandro and Paternostro, Mauro},
  journal={N. J. Phys.},
  volume={22},
  number={4},
  pages={045001},
  year={2020},
  publisher={IOP Publishing},
  doi = {10.1088/1367-2630/ab783d}
}

@article{krenn2020computer,
  title={Computer-inspired quantum experiments},
  author={Krenn, Mario and Erhard, Manuel and Zeilinger, Anton},
  journal={Nat. Rev. Phys.},
  volume={2},
  number={11},
  pages={649--661},
  year={2020},
  publisher={Nature Publishing Group UK London}, 
  doi = {10.1038/s42254-020-0230-4}
}

@article{fosel2018reinforcement,
  title={Reinforcement learning with neural networks for quantum feedback},
  author={F{\"o}sel, Thomas and Tighineanu, Petru and Weiss, Talitha and Marquardt, Florian},
  journal={Phys. Rev. X},
  volume={8},
  number={3},
  pages={031084},
  year={2018},
  publisher={APS}, 
  doi = {10.1103/PhysRevX.8.031084}
}

@article{torlai2017neural,
  title={Neural decoder for topological codes},
  author={Torlai, Giacomo and Melko, Roger G},
  journal={Phys. Rev. Lett.},
  volume={119},
  number={3},
  pages={030501},
  year={2017},
  publisher={APS},
  doi = {10.1103/PhysRevLett.119.030501}
}

@article{wittler2021integrated,
  title={Integrated tool set for control, calibration, and characterization of quantum devices applied to superconducting qubits},
  author={Wittler, Nicolas and Roy, Federico and Pack, Kevin and Werninghaus, Max and Roy, Anurag Saha and Egger, Daniel J and Filipp, Stefan and Wilhelm, Frank K and Machnes, Shai},
  journal={Phys. Rev. App.},
  volume={15},
  number={3},
  pages={034080},
  year={2021},
  publisher={APS},
  doi = {10.1103/PhysRevApplied.15.034080}
}

@article{bharti2020machine,
  title={Machine learning meets quantum foundations: A brief survey},
  author={Bharti, Kishor and Haug, Tobias and Vedral, Vlatko and Kwek, Leong-Chuan},
  journal={AVS Quant. Sci.},
  volume={2},
  number={3},
  year={2020},
  publisher={AIP Publishing},
  doi = {10.1116/5.0007529}
}

@article{zhang2002neural,
  title={Neural networks for classification: a survey},
  author={Zhang, Guoqiang Peter},
  journal={IEEE Trans. Syst. Man Cybern. Part C Appl. Rev.},
  volume={30},
  number={4},
  pages={451--462},
  year={2002},
  publisher={IEEE},
  doi = {10.1109/5326.897072}
}

@article{kingma2013auto,
  title={Auto-encoding variational bayes},
  author={Kingma, Diederik P and Welling, Max},
  journal={arXiv preprint arXiv:1312.6114},
  year={2013}
}

@article{carleo2019machine,
  title={Machine learning and the physical sciences},
  author={Carleo, Giuseppe and Cirac, Ignacio and Cranmer, Kyle and Daudet, Laurent and Schuld, Maria and Tishby, Naftali and Vogt-Maranto, Leslie and Zdeborov{\'a}, Lenka},
  journal={Rev. Mod. Phys.},
  volume={91},
  number={4},
  pages={045002},
  year={2019},
  publisher={APS},
  doi = {10.1103/RevModPhys.91.045002}
}

@article{shlomi2020graph,
  title={Graph neural networks in particle physics},
  author={Shlomi, Jonathan and Battaglia, Peter and Vlimant, Jean-Roch},
  journal={Mach. Learn.: Sci. Technol.},
  volume={2},
  number={2},
  pages={021001},
  year={2020},
  publisher={IOP Publishing},
  doi = {10.1088/2632-2153/abbf9a}
}

@article{aragon2019classifying,
  title={Classifying the large-scale structure of the universe with deep neural networks},
  author={Aragon-Calvo, Miguel A},
  journal={Mon. Not. R. Astron. Soc.},
  volume={484},
  number={4},
  pages={5771--5784},
  year={2019},
  publisher={Oxford University Press},
  doi = {10.1093/mnras/stz393}
}

@article{carrasquilla2020machine,
  title={Machine learning for quantum matter},
  author={Carrasquilla, Juan},
  journal={Adv. Phys.: X},
  volume={5},
  number={1},
  pages={1797528},
  year={2020},
  publisher={Taylor \& Francis},
  doi = {10.1080/23746149.2020.1797528}
}

@article{iten2020discovering,
  title={Discovering physical concepts with neural networks},
  author={Iten, Raban and Metger, Tony and Wilming, Henrik and Del Rio, L{\'\i}dia and Renner, Renato},
  journal={Phys. Rev. Lett.},
  volume={124},
  number={1},
  pages={010508},
  year={2020},
  publisher={APS},
  doi = {10.1103/PhysRevLett.124.010508}
}

@article{WernerPhysRevA.40.4277,
  title   = {Quantum states with {Einstein}-{Podolsky}-{Rosen} correlations admitting a hidden-variable model},
  author  = {Werner, Reinhard F.},
  journal = {Phys. Rev. A},
  volume  = {40},
  number  = {8},
  pages   = {4277--4281},
  year    = {1989},
  month   = oct,
  doi     = {10.1103/PhysRevA.40.4277},
}

@article{vempati2021witnessing,
  title   = {Witnessing negative conditional entropy},
  author  = {Vempati, Mahathi and Ganguly, Nirman and Chakrabarty, Indranil and Pati, Arun K.},
  journal = {Phys. Rev. A},
  volume  = {104},
  number  = {1},
  pages   = {012417},
  year    = {2021},
  doi     = {10.1103/PhysRevA.104.012417},
}

@article{srinidhi2024quantum,
  title   = {Quantum channels that destroy negative conditional entropy},
  author  = {Srinidhi, P. V. and Chakrabarty, Indranil and Bhattacharya, Samyadeb and Ganguly, Nirman},
  journal = {Phys. Rev. A},
  volume  = {110},
  number  = {4},
  pages   = {042423},
  year    = {2024},
  doi     = {10.1103/PhysRevA.110.042423},
}

@article{roy2025teleportation,
  title={Teleportation Fidelity of Binary Tree Quantum Repeater Networks},
  author={Roy, Soumit and Miraj, Md Rahil and Hens, Chittaranjan and Mylavarapu, Ganesh and Ghosh, Subrata and Chakrabarty, Indranil},
  journal={arXiv:2508.10417},
  year={2025},
  doi={10.48550/arXiv.2508.10417},
}

@article{patro2017non,
  title   = {Non-negativity of conditional von {N}eumann entropy and global unitary operations},
  author  = {Patro, Subhasree and Chakrabarty, Indranil and Ganguly, Nirman},
  journal = {Phys. Rev. A},
  volume  = {96},
  number  = {6},
  pages   = {062102},
  year    = {2017},
  doi     = {10.1103/PhysRevA.96.062102},
}

@article{vempati2022unital,
  title   = {A-unital operations and quantum conditional entropy},
  author  = {Vempati, Mahathi and Shah, Saumya and Ganguly, Nirman and Chakrabarty, Indranil},
  journal = {Quantum},
  volume  = {6},
  pages   = {641},
  year    = {2022},
  doi     = {10.22331/q-2022-02-02-641},
}

@article{ganguly2011entanglement,
  title   = {Entanglement witness operator for quantum teleportation},
  author  = {Ganguly, Nirman and Adhikari, Satyabrata and Majumdar, A. S. and Chatterjee, Jyotishman},
  journal = {Phys. Rev. Lett.},
  volume  = {107},
  number  = {27},
  pages   = {270501},
  year    = {2011},
  doi     = {10.1103/PhysRevLett.107.270501},
}

@article{bennett1993teleporting,
  title   = {Teleporting an unknown quantum state via dual classical and {Einstein}-{Podolsky}-{Rosen} channels},
  author  = {Bennett, Charles H. and Brassard, Gilles and Crépeau, Claude and Jozsa, Richard and Peres, Asher and Wootters, William K.},
  journal = {Phys. Rev. Lett.},
  volume  = {70},
  number  = {13},
  pages   = {1895--1899},
  year    = {1993},
  month   = mar,
  doi     = {10.1103/PhysRevLett.70.1895},
}

@article{horodecki1996teleportation,
  title   = {Teleportation, {Bell}'s inequalities and inseparability},
  author  = {Horodecki, Ryszard and Horodecki, Michał and Horodecki, Paweł},
  journal = {Phys. Lett. A},
  volume  = {222},
  number  = {1-2},
  pages   = {21--25},
  year    = {1996},
  month   = oct,
  doi     = {10.1016/0375-9601(96)00639-1},
}

@article{singh2024controlled,
  title   = {Controlled state reconstruction and quantum secret sharing},
  author  = {Singh, Pahulpreet and Chakrabarty, Indranil},
  journal = {Phys. Rev. A},
  volume  = {109},
  number  = {3},
  pages   = {032406},
  year    = {2024},
  doi     = {10.1103/PhysRevA.109.032406},
}

@article{bennett1992communication,
  title   = {Communication via one-and two-particle operators on Einstein-Podolsky-Rosen states},
  author  = {Bennett, Charles H and Wiesner, Stephen J},
  journal = {Phys. Rev. Lett.},
  volume  = {69},
  number  = {20},
  pages   = {2881},
  year    = {1992},
  doi     = {10.1103/PhysRevLett.69.2881},
}

@article{wehner2018quantum,
  title     = {Quantum internet: A vision for the road ahead},
  author    = {Wehner, Stephanie and Elkouss, David and Hanson, Ronald},
  journal   = {Science},
  volume    = {362},
  number    = {6412},
  pages     = {eaam9288},
  year      = {2018},
  publisher = {AAAS},
  doi       = {10.1126/science.aam9288},
}

@article{mundra2019broadcasting,
  title   = {Broadcasting of quantum correlations in qubit-qudit systems},
  author  = {Mundra, Rounak and Patel, Dhrumil and Chakrabarty, Indranil and Ganguly, Nirman and Chatterjee, Sourav},
  journal = {Phys. Rev. A},
  volume  = {100},
  number  = {4},
  pages   = {042319},
  year    = {2019},
  doi     = {10.1103/PhysRevA.100.042319},
}

@article{chatterjee2016broadcasting,
  title   = {Broadcasting of quantum correlations: Possibilities and impossibilities},
  author  = {Chatterjee, Sourav and Sazim, Sk and Chakrabarty, Indranil},
  journal = {Phys. Rev. A},
  volume  = {93},
  number  = {4},
  pages   = {042309},
  year    = {2016},
  doi     = {10.1103/PhysRevA.93.042309},
}

@article{mylavarapu2024teleportation,
  title   = {Teleportation fidelity of quantum repeater networks},
  author  = {Mylavarapu, Ganesh and Ghosh, Subrata and Hens, Chittaranjan and Chakrabarty, Indranil and Mitra, Subhadip},
  journal = {arXiv:2409.20304},
  year    = {2024},
  doi     = {10.48550/arXiv.2409.20304},
}

@article{jain2019asymmetric,
  title   = {Asymmetric broadcasting of quantum correlations},
  author  = {Jain, Aditya and Chakrabarty, Indranil and Chatterjee, Sourav},
  journal = {Phys. Rev. A},
  volume  = {99},
  number  = {2},
  pages   = {022315},
  year    = {2019},
  doi     = {10.1103/PhysRevA.99.022315},
}

@article{sharma2017broadcasting,
  title   = {Broadcasting quantum coherence via cloning},
  author  = {Sharma, Udit Kamal and Chakrabarty, Indranil and Shukla, Manish Kumar},
  journal = {Phys. Rev. A},
  volume  = {96},
  number  = {5},
  pages   = {052319},
  year    = {2017},
  doi     = {10.1103/PhysRevA.96.052319},
}

@article{patel2024broadcasting,
  title   = {Broadcasting of non-locality},
  author  = {Patel, Dhrumil and Roy, Arup and Chakrabarty, Indranil and Ganguly, Nirman},
  journal = {Pramana},
  volume  = {98},
  number  = {2},
  pages   = {1--11},
  year    = {2024},
  doi     = {10.1007/s12043-024-02726-1},
}

@article{mundra2021broadcasting,
  title   = {Broadcasting of NPT entanglement in two qutrit systems},
  author  = {Mundra, Rounak and Chattopadhyay, Sabuj and Chakrabarty, Indranil and Ganguly, Nirman},
  journal = {Pramana},
  volume  = {95},
  number  = {2},
  pages   = {60},
  year    = {2021},
  doi     = {10.1007/s12043-021-02098-w},
}

@article{modi2010unified,
  title   = {Unified view of quantum and classical correlations},
  author  = {Modi, Kavan and Paterek, Tomasz and Son, Wonmin and Vedral, Vlatko and Williamson, Mark},
  journal = {Phys. Rev. Lett.},
  volume  = {104},
  number  = {8},
  pages   = {080501},
  year    = {2010},
  doi     = {10.1103/PhysRevLett.104.080501},
}

@article{mylavarapu2023entanglement,
  title   = {Entanglement and Teleportation in a 1-D Network with Repeaters},
  author  = {Mylavarapu, Ganesh and Chakrabarty, Indranil and Mukherjee, Kaushiki and Huang, Minyi and Wu, Junde},
  journal = {arXiv:2306.01406},
  year    = {2023},
  doi     = {10.48550/arXiv.2306.01406},
}

@article{barnum2007generalized,
  title   = {Generalized no-broadcasting theorem},
  author  = {Barnum, Howard and Barrett, Jonathan and Leifer, Matthew and Wilce, Alexander},
  journal = {Phys. Rev. Lett.},
  volume  = {99},
  number  = {24},
  pages   = {240501},
  year    = {2007},
  doi     = {10.1103/PhysRevLett.99.240501},
}

@article{wootters1982single,
  title     = {A single quantum cannot be cloned},
  author    = {Wootters, William K and Zurek, Wojciech H},
  journal   = {Nature},
  volume    = {299},
  number    = {5886},
  pages     = {802--803},
  year      = {1982},
  publisher = {Nature Publishing Group},
  doi       = {10.1038/299802a0},
}

@article{buzek1997broadcasting,
  title   = {Broadcasting of entanglement via local copying},
  author  = {Buzek, Vladimir and Vedral, Vlatko and Plenio, Martin B. and Knight, Peter L. and Hillery, Mark},
  journal = {Phys. Rev. A},
  volume  = {55},
  number  = {5},
  pages   = {3327},
  year    = {1997},
  doi     = {10.1103/PhysRevA.55.3327},
}

@article{adhikari2008broadcasting,
  title   = {Broadcasting of continuous-variable entanglement},
  author  = {Adhikari, Satyabrata and Majumdar, Archan S. and Nayak, Nilakantha},
  journal = {Phys. Rev. A},
  volume  = {77},
  number  = {4},
  pages   = {042301},
  year    = {2008},
  doi     = {10.1103/PhysRevA.77.042301},
}

@article{horodecki2001distillation,
  title   = {Distillation and bound entanglement},
  author  = {Horodecki, Pawel and Horodecki, Ryszard},
  journal = {Quantum Inf. Comput.},
  volume  = {1},
  number  = {1},
  pages   = {45--75},
  year    = {2001},
  doi     = {10.26421/QIC1.1-4},
}

@article{chakrabarty2010teleportation,
  title   = {Teleportation via a mixture of a two qubit subsystem of a N-qubit W and GHZ state},
  author  = {Chakrabarty, Indranil},
  journal = {Eur. Phys. J. D},
  volume  = {57},
  number  = {2},
  pages   = {265--269},
  year    = {2010},
  doi     = {10.1140/epjd/e2010-00014-6},
}

@article{chakrabarty2011deletion,
  title   = {Deletion, Bell’s inequality, teleportation},
  author  = {Chakrabarty, Indranil and Ganguly, Nirman and Choudhury, Binayak S},
  journal = {Quantum Inf. Process.},
  volume  = {10},
  number  = {1},
  pages   = {27--32},
  year    = {2011},
  doi     = {10.1007/s11128-010-0167-0},
}

@article{adhikari2008quantum,
  title   = {Quantum cloning, Bell's inequality and teleportation},
  author  = {Adhikari, Satyabrata and Ganguly, Nirman and Chakrabarty, Indranil and Choudhury, Binayak S.},
  journal = {J. Phys. A: Math. Theor.},
  volume  = {41},
  number  = {41},
  pages   = {415302},
  year    = {2008},
  doi     = {10.1088/1751-8113/41/41/415302},
}

@article{Ahmed2021qstcgan,
  title   = {Quantum State Tomography with Conditional Generative Adversarial Networks},
  author  = {Ahmed, Shahnawaz and Sánchez Muñoz, Carlos and Nori, Franco and Kockum, Anton Frisk},
  journal = {Phys. Rev. Lett.},
  volume  = {127},
  number  = {14},
  pages   = {140502},
  year    = {2021},
  doi     = {10.1103/PhysRevLett.127.140502},
}

@article{Durall2020CombatingMC,
  title   = {Combating Mode Collapse in GAN training: An Empirical Analysis using Hessian Eigenvalues},
  author  = {Ricard Durall and Avraam Chatzimichailidis and Peter Labus and Janis Keuper},
  journal = {arXiv:2012.09673},
  year    = {2020},
  doi     = {10.48550/arXiv.2012.09673},
}

@article{Koutn2022neuralnetworkqst,
  title   = {Neural-network quantum state tomography},
  author  = {Koutný, Dominik and Motka, Libor and Hradil, Zdeněk and Řeháček, Jaroslav and Sánchez-Soto, Luis L.},
  journal = {Phys. Rev. A},
  volume  = {106},
  number  = {1},
  pages   = {012409},
  year    = {2022},
  doi     = {10.1103/PhysRevA.106.012409},
}

@article{Goodfellow2020gan,
  author  = {Goodfellow, Ian and Pouget-Abadie, Jean and Mirza, Mehdi and Xu, Bing and Warde-Farley, David and Ozair, Sherjil and Courville, Aaron and Bengio, Yoshua},
  title   = {Generative adversarial networks},
  year    = {2020},
  journal = {Commun. ACM},
  volume  = {63},
  number  = {11},
  pages   = {139–144},
  month   = oct,
  doi     = {10.1145/3422622},
}

@article{Lloyd2018qgal,
  title   = {Quantum Generative Adversarial Learning},
  author  = {Lloyd, Seth and Weedbrook, Christian},
  journal = {Phys. Rev. Lett.},
  volume  = {121},
  number  = {4},
  pages   = {040502},
  year    = {2018},
  doi     = {10.1103/PhysRevLett.121.040502},
}

@article{Horodecki1998,
  author  = {Horodecki, Michal and Horodecki, Pawel and Horodecki, Ryszard},
  title   = {Mixed-State Entanglement and Distillation: Is There a "Bound" Entanglement in Nature?},
  journal = {Phys. Rev. Lett.},
  volume  = {80},
  number  = {24},
  pages   = {5239--5242},
  year    = {1998},
  doi     = {10.1103/PhysRevLett.80.5239}
}

@article{Chitambar2019,
  author    = {Chitambar, Eric and Gour, Gilad},
  title     = {Quantum resource theories},
  journal   = {Rev. Mod. Phys.},
  volume    = {91},
  number    = {2},
  pages     = {025001},
  year      = {2019},
  month     = {Apr},
  doi       = {10.1103/RevModPhys.91.025001},
}

@inproceedings{Goodfellow2014,
  author    = {Goodfellow, Ian J. and Pouget-Abadie, Jean and Mirza, Mehdi and Xu, Bing
               and Warde-Farley, David and Ozair, Sherjil and Courville, Aaron and Bengio, Yoshua},
  title     = {Generative Adversarial Nets},
  booktitle = {Advances in Neural Information Processing Systems},
  editor    = {Ghahramani, Zoubin and Welling, Max and Cortes, Corinna and Lawrence, Neil D. and Weinberger, Kilian Q.},
  volume    = {27},
  pages     = {2672--2680},
  year      = {2014},
  publisher = {Curran Associates, Inc.},
  doi       = {10.48550/arXiv.1406.2661},
  url       = {https://arxiv.org/abs/1406.2661}
}

@article{mirza2014conditionalgenerativeadversarialnets,
  author  = {Mirza, Mehdi and Osindero, Simon},
  title   = {Conditional Generative Adversarial Nets},
  journal = {arXiv preprint arXiv:1411.1784},
  year    = {2014},
  doi     = {10.48550/arXiv.1411.1784},
  url     = {https://arxiv.org/abs/1411.1784}
}

@article{molesky2018inverse,
  title   = {Inverse design in nanophotonics},
  author  = {Molesky, Sean and Lin, Zin and Piggott, Alexander Y. and Jin, Weiliang and Vuckovic, Jelena and Rodriguez, Alejandro W.},
  journal = {Nat. Photon.},
  volume  = {12},
  number  = {11},
  pages   = {659--670},
  year    = {2018},
  doi     = {10.1038/s41566-018-0246-9},
}

@article{sanchez2018inverse,
  title   = {Inverse molecular design using machine learning: Generative models for matter engineering},
  author  = {Sanchez-Lengeling, Benjamin and Aspuru-Guzik, Alan},
  journal = {Science},
  volume  = {361},
  number  = {6400},
  pages   = {360--365},
  year    = {2018},
  doi     = {10.1126/science.aat2663},
}

@article{malkiel2018plasmonic,
  title   = {Plasmonic nanostructure design and characterization via deep learning},
  author  = {Malkiel, Itzik and Mrejen, Michael and Nagler, Achiya and Arieli, Uri and Wolf, Lior and Suchowski, Haim},
  journal = {Light Sci. Appl.},
  volume  = {7},
  number  = {1},
  pages   = {60},
  year    = {2018},
  doi     = {10.1038/s41377-018-0060-7},
}

@book{joannopoulos2008photonic,
  title     = {Photonic Crystals: Molding the Flow of Light},
  author    = {Joannopoulos, John D. and Johnson, Steven G. and Winn, Joshua N. and Meade, Robert D.},
  publisher = {Princeton University Press},
  year      = {2008},
  edition   = {2nd},
  doi       = {10.2307/j.ctvcm4gz9},
}

@article{smith2004metamaterials,
  title   = {Metamaterials and negative refractive index},
  author  = {Smith, David R and Pendry, John B and Wiltshire, Mike CK},
  journal = {Science},
  volume  = {305},
  number  = {5685},
  pages   = {788--792},
  year    = {2004},
  doi     = {10.1126/science.1096796},
}

@article{schmidt2019recent,
  title   = {Recent advances and applications of machine learning in solid-state materials science},
  author  = {Schmidt, Jonathan and Marques, Mário RG and Botti, Silvana and Marques, Miguel AL},
  journal = {npj Comput. Mater.},
  volume  = {5},
  number  = {1},
  pages   = {83},
  year    = {2019},
  doi     = {10.1038/s41524-019-0221-0},
}

@article{zhao2019acoustic,
  title   = {Intelligent acoustic material design with deep learning},
  author  = {Zhao, Shuyu and Lee, Gunhee and Li, Jensen and Hu, Xiaoshi},
  journal = {APL Mater.},
  volume  = {7},
  number  = {10},
  pages   = {101103},
  year    = {2019},
  doi     = {10.1063/1.5124453},
}

@article{hu2017discovering,
  title   = {Discovering phases, phase transitions, and crossovers through unsupervised machine learning: A critical examination},
  author  = {Hu, Wenjian and Singh, Rajiv RP and Scalettar, Richard T},
  journal = {Phys. Rev. E},
  volume  = {95},
  number  = {6},
  pages   = {062122},
  year    = {2017},
  doi     = {10.1103/PhysRevE.95.062122},
}

@article{paganini2018accelerating,
  title   = {Accelerating science with generative adversarial networks: an application to 3D particle showers in multilayer calorimeters},
  author  = {Paganini, Michela and de Oliveira, Luke and Nachman, Benjamin},
  journal = {Phys. Rev. Lett.},
  volume  = {120},
  number  = {4},
  pages   = {042003},
  year    = {2018},
  doi     = {10.1103/PhysRevLett.120.042003},
}

@article{Hu2019qgan,
  title   = {Quantum generative adversarial learning in a superconducting quantum circuit},
  author  = {Hu, Ling and Wu, Shu-Hao and Cai, Weizhou and Ma, Yuwei and Mu, Xianghao and Xu, Yuan and Wang, Haiyan and Song, Yipu and Deng, Dong-Ling and Zou, Chang-Ling and Sun, Luyan},
  journal = {Sci. Adv.},
  volume  = {5},
  number  = {1},
  pages   = {eaav2761},
  year    = {2019},
  doi     = {10.1126/sciadv.aav2761},
}

@article{horodecki2009quantum,
  title   = {Quantum entanglement},
  author  = {Horodecki, Ryszard and Horodecki, Pawel and Horodecki, Michal and Horodecki, Karol},
  journal = {Rev. Mod. Phys.},
  volume  = {81},
  number  = {2},
  pages   = {865--942},
  year    = {2009},
  doi     = {10.1103/RevModPhys.81.865},
}

@article{giovannetti2011advances,
  title   = {Advances in quantum metrology},
  author  = {Giovannetti, Vittorio and Lloyd, Seth and Maccone, Lorenzo},
  journal = {Nat. Photon.},
  volume  = {5},
  number  = {4},
  pages   = {222--229},
  year    = {2011},
  doi     = {10.1038/nphoton.2011.35},
}

@article{terhal2015quantum,
  title   = {Quantum error correction for quantum memories},
  author  = {Terhal, Barbara M},
  journal = {Rev. Mod. Phys.},
  volume  = {87},
  number  = {2},
  pages   = {307},
  year    = {2015},
  doi     = {10.1103/RevModPhys.87.307},
}

@article{coopmans2021netsquid,
  title   = {{NetSquid}, a {NET}work {S}imulator for {QU}antum {I}nformation using {D}iscrete events},
  author  = {Coopmans, Tim and Knegjens, Robert and Dahlberg, Axel and Maier, David and Nijsten, Loek and de Oliveira Filho, Julio and Papendrecht, Martijn and Rabbie, Julian and Rozpedek, Filip and Skrzypczyk, Matthew and Wubben, Leon and de Jong, Walter and Podber{\v{s}}ek, Damian and Torres-Knoop, Ariana and Elkouss, David and Wehner, Stephanie},
  journal = {Commun. Phys.},
  volume  = {4},
  pages   = {164},
  year    = {2021},
  doi     = {10.1038/s42005-021-00647-8},
}

@inproceedings{cohen2019gauge,
  title     = {Gauge Equivariant Convolutional Networks and the Icosahedral {CNN}},
  author    = {Cohen, Taco S. and Weiler, Maurice and Kicanaoglu, Berkay and Welling, Max},
  booktitle = {Proceedings of the 36th International Conference on Machine Learning},
  series    = {PMLR},
  volume    = {97},
  pages     = {1321--1330},
  year      = {2019},
}

@inproceedings{ho2020diffusion,
  author    = {Ho, Jonathan and Jain, Ajay and Abbeel, Pieter},
  title     = {Denoising Diffusion Probabilistic Models},
  booktitle = {Advances in Neural Information Processing Systems},
  volume    = {33},
  pages     = {6840--6851},
  year      = {2020},
  doi       = {10.48550/arXiv.2006.11239},
  url       = {https://arxiv.org/abs/2006.11239}
}

@inproceedings{heusel2018ganstrainedtimescaleupdate,
author = {Heusel, Martin and Ramsauer, Hubert and Unterthiner, Thomas and Nessler, Bernhard and Hochreiter, Sepp},
title = {GANs trained by a two time-scale update rule converge to a local nash equilibrium},
year = {2017},
isbn = {9781510860964},
publisher = {Curran Associates Inc.},
address = {Red Hook, NY, USA},
booktitle = {Proceedings of the 31st International Conference on Neural Information Processing Systems},
pages = {6629–6640},
numpages = {12},
location = {Long Beach, California, USA},
series = {NIPS'17}
}
\appendix

\renewcommand{\thefigure}{\thesection\arabic{figure}}
\renewcommand{\thetable}{\thesection\arabic{table}}
\renewcommand{\theequation}{\thesection\arabic{equation}}

\makeatletter
\@addtoreset{figure}{section}
\@addtoreset{table}{section}
\@addtoreset{equation}{section}
\makeatother

\section{Reproducibility Details}
\label{app:reproducibility}
This section contains all necessary information for the full reproduction of the results presented in the main text.

\subsection{Model Architectures}

All generator models take a 100-dimensional latent vector $z \sim \mathcal{N}(0, I)$ and include a residual connection for stability. A single discriminator architecture is shared across experiments.

\paragraph{Generator models.}
\text{Cholesky Decomposition:} Parameterizes $\rho = LL^\dagger$, where any matrix of the form $LL^\dagger$ is inherently positive semi-definite, enforcing this constraint by construction. Uses a multilayer perceptron (MLP) with Leaky ReLU activations, LayerNorm, and a trace normalization step.

\text{Enhanced LDL Decomposition:} Uses $\rho = LDL^\dagger$ with $L$ lower triangular (unit diagonal) and $D$ diagonal with strictly positive entries ($\epsilon = 10^{-6}$ regularization). Ensures Hermiticity by construction and trace-one via normalization.

\text{Direct Matrix Generator:} Outputs a 32-dimensional vector reshaped into a $4 \times 4$ Hermitian matrix. Physical constraints (Hermiticity, PSD, trace-one) are enforced entirely through explicit loss functions rather than architectural design, serving as a baseline to demonstrate the benefit of constraint embedding.

\paragraph{Discriminator.} An MLP classifier for flattened $4 \times 4$ matrices (32 real values): 1. Input: Linear(32 $\to$ 256) + LeakyReLU + LayerNorm + Dropout. 2. Hidden Layers: 256 $\to$ 512, 512 $\to$ 256, 256 $\to$ 128.3. Output: Linear(128 $\to$ 1) with Sigmoid





\subsection{Training Hyperparameters}

All models were trained with the RMSprop optimizer. Loss weights were tuned empirically to balance contributions.

\begin{table}[h]
    \centering
    \caption{Training hyperparameters and loss weights for decomposition-based generators (Models 1 \& 2).}
    \begin{tabular}{ll}
        \toprule
        Hyperparameter & Value \\
        \midrule
        Optimizer & RMSprop \\
        Learning Rate & $10^{-5}$ \\
        Batch Size & 512 \\
        Training Epochs & 10,000 \\
        \midrule
        $\lambda_{psd}$ (PSD Loss) & 10.0 \\
        $\lambda_{trace}$ (Trace Loss) & 10.0 \\
        $\lambda_{herm}$ (Hermiticity Loss, Model 3 only) & 5.0 \\
        $\lambda_{task}$ (Task-Specific Loss) & $5.0 \times m_{\text{task}}$ \\
        $\lambda_{diversity}$ (Diversity Loss) & 0.5 \\
        \bottomrule
    \end{tabular}
\end{table}

\noindent
The task-specific multiplier $m_{\text{task}}$ accounts for varying optimization complexity:
\text{Teleportation:} $m_{\text{task}} = 1.0$ (simpler $F_{\max} > 2/3$ criterion). \text{Nonlocal Broadcasting:} $m_{\text{task}} = 1.2$. \text{Local Broadcasting:} $m_{\text{task}} = 1.5$ (most complex Positive Partial Transpose (PPT) conditions).\\

\paragraph{Rationale.} For decomposition-based models (Cholesky/LDL), physics constraints (PSD, trace-one) are architecturally enforced, reducing the need for strong penalty terms. The moderate physics weights provide soft regularization while the elevated task-specific weight ensures the optimization focuses on the actual objective. The diversity term encourages exploration without dominating convergence. RMSprop was selected as the default optimizer after comparing Adadelta, AdamW, Adam, and RMSprop; it showed greater stability on constrained broadcasting tasks.

\subsection{Dataset Generation}
The training datasets used for the discriminator were generated as follows:\\
1. \emph{Werner-like states:} Training data were generated by uniformly sampling the parameters \(p \in [0,1]\) and \(\alpha, \beta\) from a uniform distribution on the unit circle (subject to \(\alpha^2 + \beta^2 = 1\)), retaining only states satisfying the task-specific criterion (e.g., \(p(1+4\alpha\beta) > 1\) for teleportation; see Appendix~\ref{app:teleportation_ppt_equivalence}). For each task, 2000 unique states were generated.\\
2. \emph{Bell-diagonal states:} Training data were generated by uniformly sampling the parameters \(c_1, c_2, c_3\) from the range \([-1, 1]\), subject to the constraints required for the state to be a valid resource for the specified task (e.g., \(|c_1|+|c_2|+|c_3|>1\) for teleportation). For each task, 2000 unique states were generated.

\section{Loss Function Formulation}
\label{app:loss}
This section provides the explicit mathematical forms of the loss components referenced in Eq.~\eqref{eq:generator_loss} of the main text. For a batch of $N$ generated density matrices $\{\rho_{\text{gen}}^{(i)}\}_{i=1}^N$, the individual components are:

\begin{itemize}
    \item \emph{Adversarial loss:}
    $$ \mathcal{L}_{\text{adv}} = -\frac{1}{N} \sum_{i=1}^{N} \log(D(\rho_{\text{gen}}^{(i)})) $$

    \item \emph{Trace loss:}
    $$ \mathcal{L}_{\text{trace}} = \frac{1}{N} \sum_{i=1}^{N} \left| \text{Tr}(\rho_{\text{gen}}^{(i)}) - 1 \right| $$

    \item \emph{PSD loss:} For eigenvalues $\{\lambda_j^{(i)}\}$ of each $\rho_{\text{gen}}^{(i)}$:
    $$ \mathcal{L}_{\text{psd}} = \frac{1}{N} \sum_{i=1}^{N} \sum_{j} \max(0, -\lambda_j^{(i)}) $$

    \item \emph{Task-specific loss} ($\mathcal{L}_{\text{task}}$): Penalizes states failing the fidelity criterion—$F_{\max} \le 2/3$ for teleportation, or PPT-based conditions for broadcasting.
\end{itemize}

\section{Experimental Results}
\label{app:results}

Tables~\ref{tab:bell_local}--\ref{tab:werner_tele} present detailed performance metrics across models, tasks, and dataset sizes. Abbreviated model names: Cholesky (Cholesky Decomposition), LDL (Enhanced LDL Decomposition), Direct (Direct Matrix Generator). Analysis of these results is provided in Section~\ref{sec:resource_states_cgan} of the main text.
\begin{table}[!ht]
\centering
\caption{Metrics for Bell Diagonal — Local Broadcasting.}
\label{tab:bell_local}
\begin{tabular}{lrlll}
\toprule
Model & Train Size & Accuracy & Fidelity & FID \\
\midrule
Cholesky & 500 & 0.936 & 0.590 & 0.517 \\
Cholesky & 1000 & 0.880 & 0.586 & 0.451 \\
Cholesky & 2000 & 0.941 & 0.581 & 0.452 \\
LDL & 500 & 0.988 & 0.483 & 0.885 \\
LDL & 1000 & 0.679 & 0.593 & 0.467 \\
LDL & 2000 & 0.652 & 0.659 & 0.560 \\
Direct & 500 & 0.100 & 0.681 & 0.678 \\
Direct & 1000 & 0.120 & 0.514 & 2.236 \\
Direct & 2000 & 0.150 & 0.467 & 2.394 \\
\bottomrule
\end{tabular}
\end{table}

\begin{table}[!ht]
\centering
\caption{Metrics for Bell Diagonal — Nonlocal Broadcasting.}
\label{tab:bell_nonlocal}
\begin{tabular}{lrlll}
\toprule
Model & Train Size & Accuracy & Fidelity & FID \\
\midrule
Cholesky & 500 & 0.881 & 0.710 & 0.083 \\
Cholesky & 1000 & 0.876 & 0.700 & 0.049 \\
Cholesky & 2000 & 0.950 & 0.701 & 0.123 \\
LDL & 500 & 0.983 & 0.678 & 0.121 \\
LDL & 1000 & 0.770 & 0.694 & 0.127 \\
LDL & 2000 & 0.982 & 0.704 & 0.148 \\
Direct & 500 & 0.120 & 0.744 & 0.384 \\
Direct & 1000 & 0.150 & 0.703 & 0.489 \\
Direct & 2000 & 0.180 & 0.702 & 0.442 \\
\bottomrule
\end{tabular}
\end{table}

\begin{table}[!ht]
\centering
\caption{Metrics for Bell Diagonal — Teleportable.}
\label{tab:bell_tele}
\begin{tabular}{lrlll}
\toprule
Model & Train Size & Accuracy & Fidelity & FID \\
\midrule
Cholesky & 500 & 0.743 & 0.790 & 0.011 \\
Cholesky & 1000 & 0.852 & 0.789 & 0.016 \\
Cholesky & 2000 & 0.979 & 0.769 & 0.011 \\
LDL & 500 & 0.980 & 0.798 & 0.010 \\
LDL & 1000 & 0.927 & 0.796 & 0.015 \\
LDL & 2000 & 0.972 & 0.787 & 0.015 \\
Direct & 500 & 0.450 & 0.792 & 0.075 \\
Direct & 1000 & 0.500 & 0.787 & 0.140 \\
Direct & 2000 & 0.550 & 0.770 & 0.145 \\
\bottomrule
\end{tabular}
\end{table}

\begin{table}[!ht]
\centering
\caption{Metrics for Werner-like — Local Broadcasting.}
\label{tab:werner_local}
\begin{tabular}{lrlll}
\toprule
Model & Train Size & Accuracy & Fidelity & FID \\
\midrule
Cholesky & 500 & 0.972 & 0.962 & 0.004 \\
Cholesky & 1000 & 0.926 & 0.961 & 0.004 \\
Cholesky & 2000 & 0.976 & 0.962 & 0.003 \\
LDL & 500 & 0.990 & 0.959 & 0.004 \\
LDL & 1000 & 0.945 & 0.959 & 0.005 \\
LDL & 2000 & 0.975 & 0.962 & 0.003 \\
Direct & 500 & 0.300 & 0.963 & 0.047 \\
Direct & 1000 & 0.320 & 0.962 & 0.041 \\
Direct & 2000 & 0.350 & 0.962 & 0.055 \\
\bottomrule
\end{tabular}
\end{table}

\begin{table}[!ht]
\centering
\caption{Metrics for Werner-like — Nonlocal Broadcasting.}
\label{tab:werner_nonlocal}
\begin{tabular}{lrlll}
\toprule
Model & Train Size & Accuracy & Fidelity & FID \\
\midrule
Cholesky & 500 & 0.945 & 0.943 & 0.014 \\
Cholesky & 1000 & 0.974 & 0.941 & 0.008 \\
Cholesky & 2000 & 0.977 & 0.940 & 0.008 \\
LDL & 500 & 0.981 & 0.940 & 0.010 \\
LDL & 1000 & 0.983 & 0.941 & 0.008 \\
LDL & 2000 & 0.983 & 0.941 & 0.010 \\
Direct & 500 & 0.400 & 0.940 & 0.068 \\
Direct & 1000 & 0.450 & 0.940 & 0.055 \\
Direct & 2000 & 0.500 & 0.941 & 0.064 \\
\bottomrule
\end{tabular}
\end{table}

\begin{table}[!ht]
\centering
\caption{Metrics for Werner-like — Teleportable.}
\label{tab:werner_tele}
\begin{tabular}{lrlll}
\toprule
Model & Train Size & Accuracy & Fidelity & FID \\
\midrule
Cholesky & 500 & 0.968 & 0.922 & 0.017 \\
Cholesky & 1000 & 0.963 & 0.907 & 0.008 \\
Cholesky & 2000 & 1.000 & 0.926 & 0.138 \\
LDL & 500 & 0.979 & 0.922 & 0.015 \\
LDL & 1000 & 0.959 & 0.907 & 0.010 \\
LDL & 2000 & 0.990 & 0.926 & 0.014 \\
Direct & 500 & 0.500 & 0.926 & 0.050 \\
Direct & 1000 & 0.550 & 0.908 & 0.046 \\
Direct & 2000 & 0.600 & 0.927 & 0.051 \\
\bottomrule
\end{tabular}
\end{table}

\vspace{4pt}
\noindent\textit{Note:} Metric definitions are provided in Section~\ref{sec:gans} of the main text.

\section{Counter Examples: Suboptimal Training Epochs}
\label{app:counter_examples}

To demonstrate the importance of proper epoch selection, we include counter examples at epoch 300 showing performance at early training stages. These illustrate degraded accuracy compared to the epoch 3000 results in Figs.~\ref{fig:theoretical_bell_diagonal_regions} and~\ref{fig:theoretical_werner_like_regions}.

\subsection{Rationale for Different Epoch Selection}

The counter example figures use different epoch numbers than the main text (epoch 3000) for methodological reasons:
\begin{enumerate}
    \item Worse metrics at epoch 3000 with the same architecture would require intentionally degraded hyperparameters, introducing confounding factors.
    \item Sampling different epochs along the same training trajectory demonstrates natural performance variation without altering the model.
    \item This shows epoch 3000 was chosen deliberately based on empirical stability~\cite{Goodfellow2014,salimans2016improved}, not arbitrarily.
\end{enumerate}
Importantly, main-text figures use the \emph{same} model weights (Model 1, epoch 3000); counterexamples explicitly vary the epoch to illustrate training dynamics.

\subsection{Counter Example Figures}

\begin{figure*}[!ht]
  \centering
  \subfloat[Local broadcasting\label{fig:counter_bell_local}]{
    \includegraphics[width=0.32\textwidth]{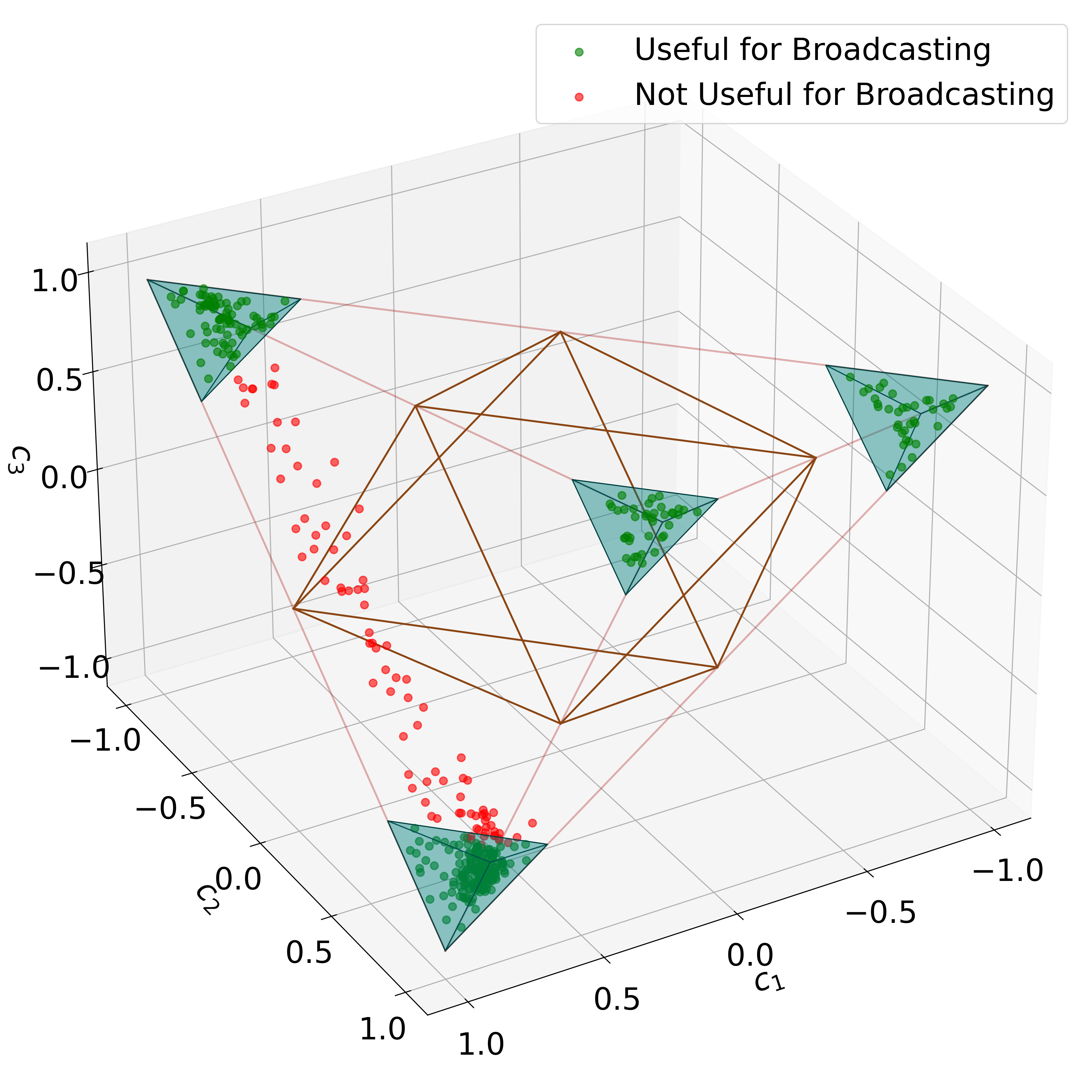}}
  \hfill
  \subfloat[Non-local broadcasting\label{fig:counter_bell_nonlocal}]{
    \includegraphics[width=0.32\textwidth]{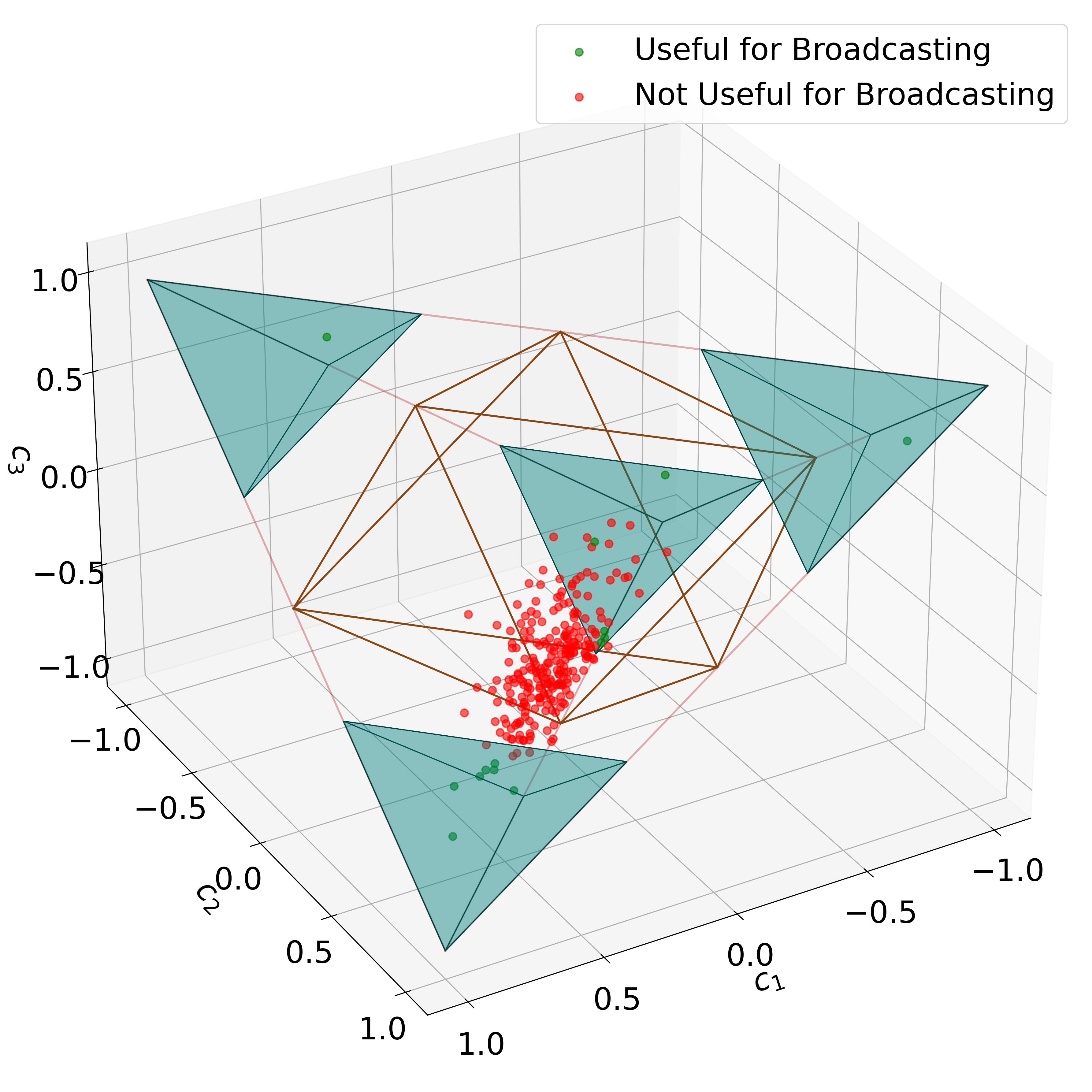}}
  \hfill
  \subfloat[Teleportation\label{fig:counter_bell_tele}]{
    \includegraphics[width=0.32\textwidth]{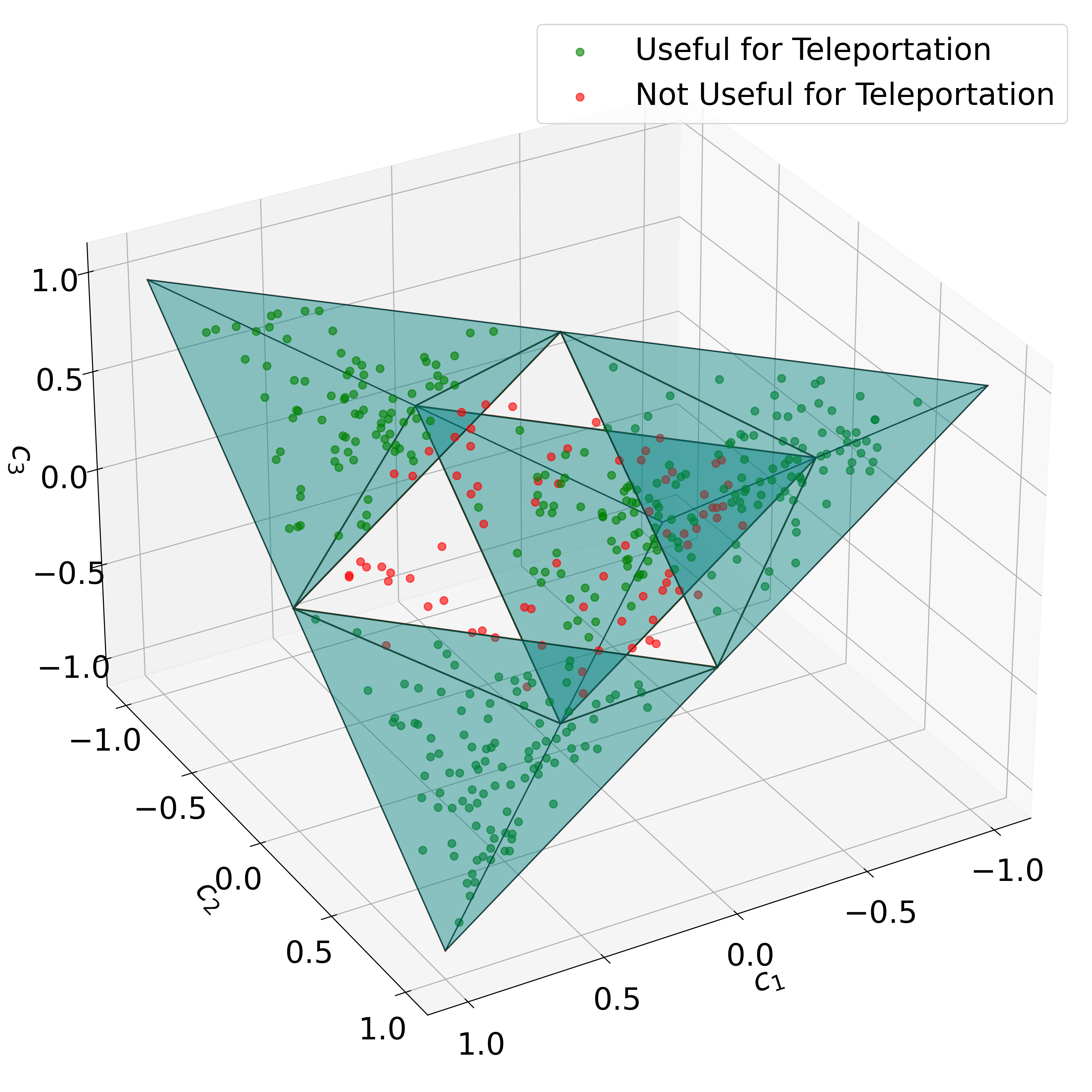}}

  \caption{Counter example: Bell-diagonal states at epoch 300 (Model 1: Cholesky, $N=2000$).
  Subplots: (1)~local broadcasting, (2)~non-local broadcasting, (3)~teleportation.
  Accuracy at epoch 300 vs. 3000: local $79\%$ vs. $95\%$, non-local $5\%$ vs. $91\%$, teleportation $82\%$ vs. $97\%$.
  Early training shows substantially degraded performance, particularly for non-local broadcasting where accuracy remains below $10\%$.}
  \label{fig:counter_bell_diagonal_all}
\end{figure*}

\begin{figure*}[!ht]
  \centering
  \subfloat[Local broadcasting\label{fig:counter_werner_local}]{
    \includegraphics[width=0.32\textwidth]{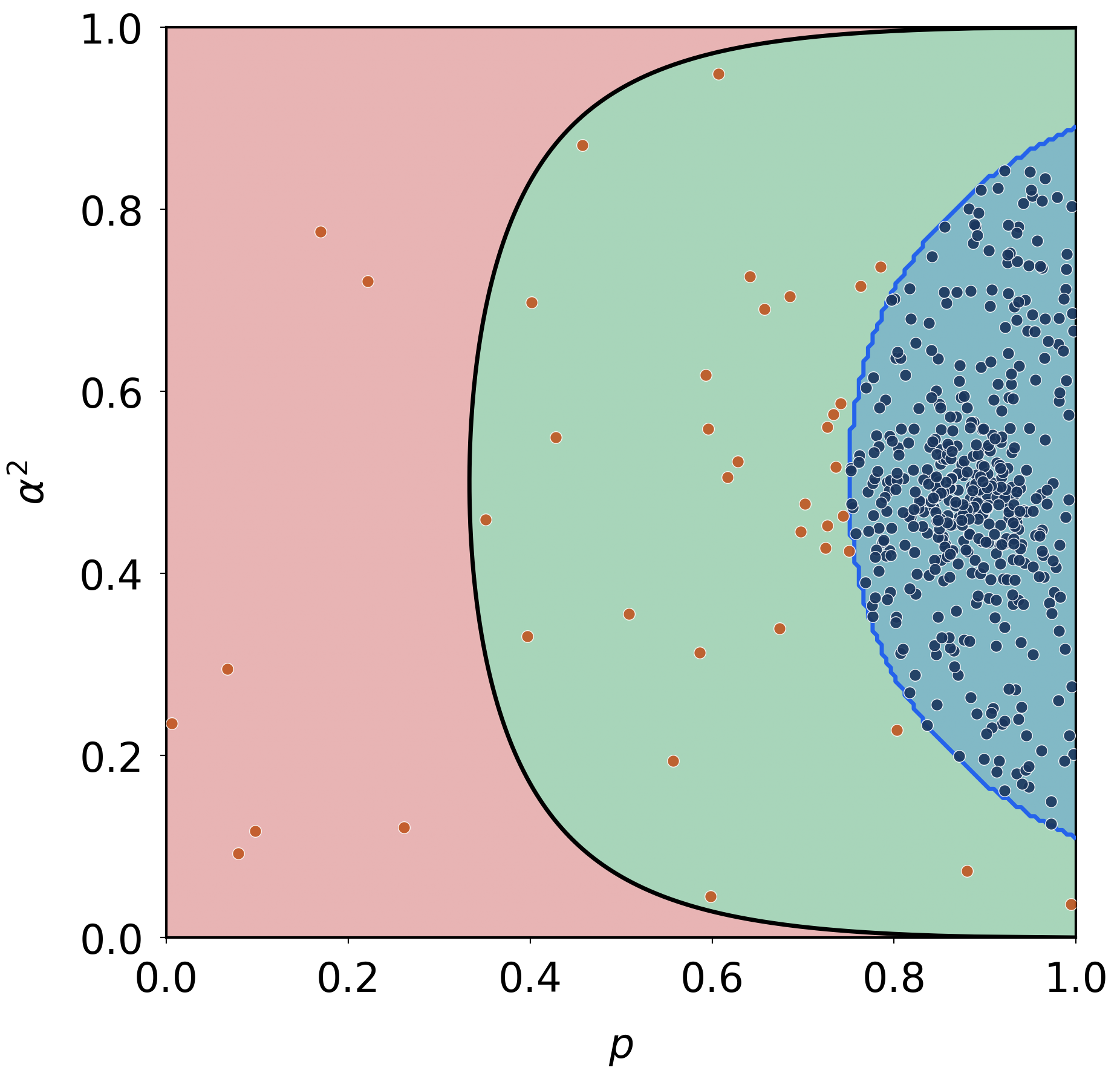}}
  \hfill
  \subfloat[Non-local broadcasting\label{fig:counter_werner_nonlocal}]{
    \includegraphics[width=0.32\textwidth]{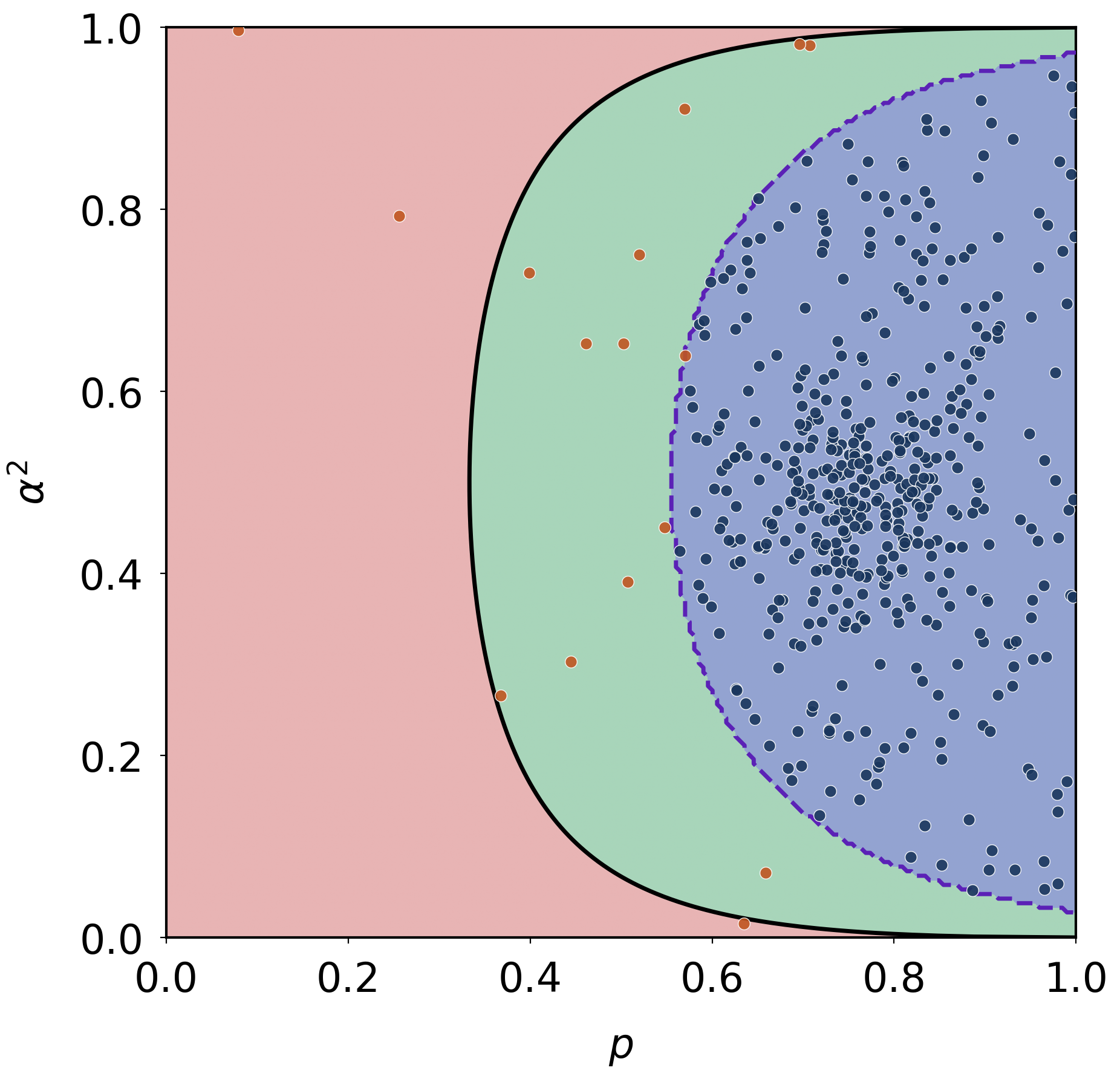}}
  \hfill
  \subfloat[Teleportation\label{fig:counter_werner_tele}]{
    \includegraphics[width=0.32\textwidth]{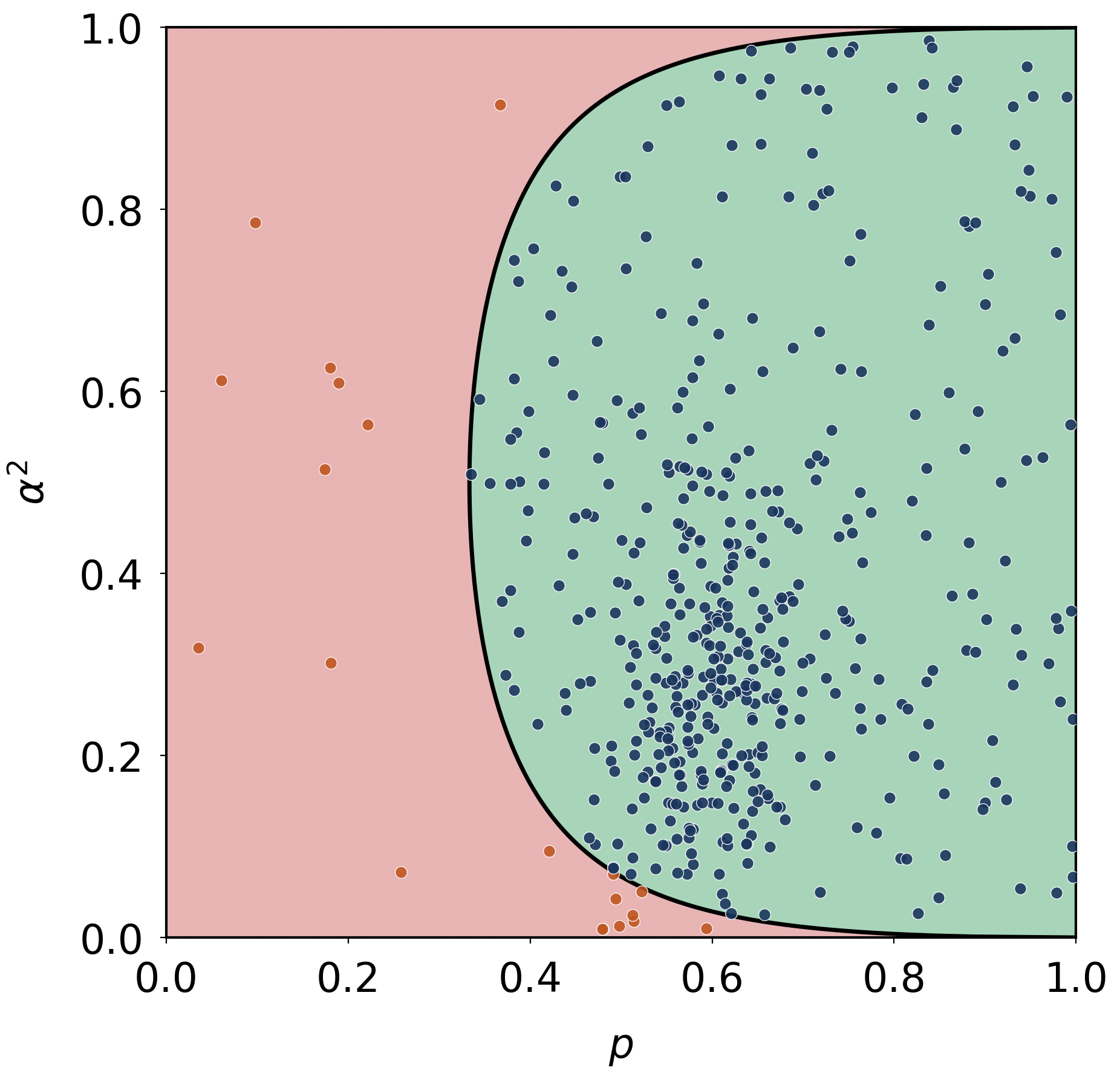}}

  \caption{Counter example: Werner-like states at epoch 300 (Model 1: Cholesky, $N=2000$).
  Subplots: (1)~local broadcasting, (2)~non-local broadcasting, (3)~teleportation.
  Accuracy at epoch 300 vs. 3000: local $92\%$ vs. $99\%$, non-local $97\%$ vs. $94\%$, teleportation $96\%$ vs. $91\%$.
  Unlike Bell-diagonal states, Werner-like generation achieves high accuracy early ($>90\%$ by epoch 300). Non-monotonic trajectories—where epoch 300 outperforms epoch 3000 for non-local/teleportation tasks—illustrate training instability rather than consistent improvement.}
  \label{fig:counter_werner_like_all}
\end{figure*}

Figures~\ref{fig:counter_bell_diagonal_all} and~\ref{fig:counter_werner_like_all} show generated states at epoch 300 (Model 1: Cholesky, $N=2000$) for all six task-state combinations. Comparing these early-training results with the epoch 3000 scatter plots in Figs.~\ref{fig:theoretical_bell_diagonal_regions} and~\ref{fig:theoretical_werner_like_regions} demonstrates the impact of epoch selection on generation quality.

\subsection*{Computational Complexity of State Generation}
\begin{figure*}[!ht]
  \centering
  \includegraphics[width=0.95\textwidth]{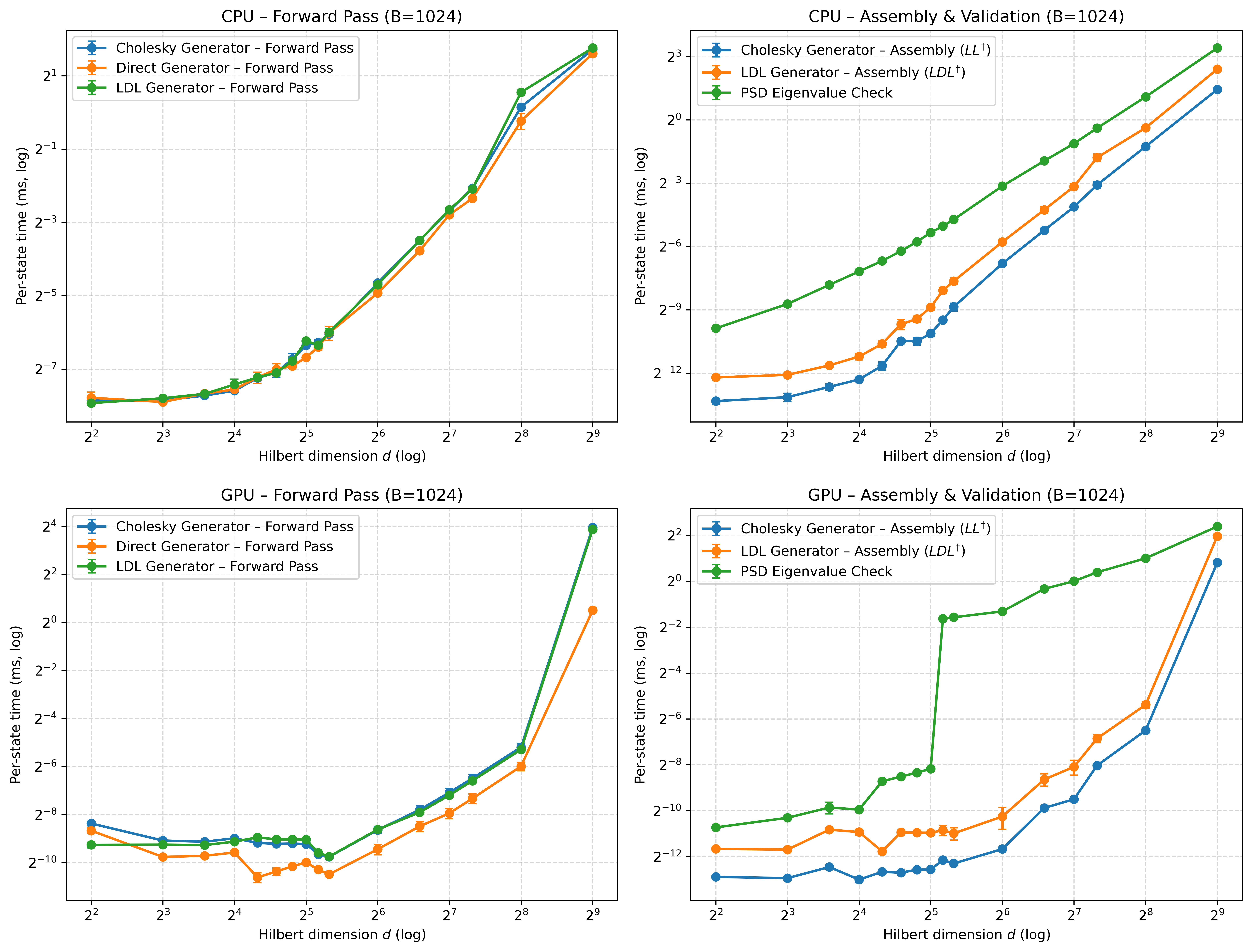}
  \caption{
Empirical scaling of per-state inference and validation costs with Hilbert dimension $d$
at batch size $B=1024$.
Top row: CPU results. Bottom row: GPU results.
Left: Forward-pass costs for Direct, Cholesky, and LDL generators, scaling $\sim d^2$.
Right: Assembly ($LL^\dagger$, $LDL^\dagger$) and PSD/PPT eigenvalue checks, scaling $\sim d^3$.
Error bars indicate 95\% confidence intervals across 10 repeats.
}
  \label{fig:scaling}
\end{figure*}

To understand the scaling behavior of our generator models and to contextualize the empirical benchmarks presented in Fig.~\ref{fig:scaling}, we performed an asymptotic analysis of the computational costs. This analysis validates the empirically observed $\Theta(d^2)$ and $\Theta(d^3)$ scaling behaviors and clarifies the trade-offs between the direct and decomposition-based approaches when extending the framework to higher Hilbert space dimensions $d$.

We analyze the computational cost of generating $K$ density matrices 
$\rho \in \mathbb{C}^{d \times d}$ during inference, where $d$ is the Hilbert space dimension 
($d=4$ for the two-qubit case considered in the main text). 
Each generator is implemented as a multilayer perceptron (MLP) with parameter count
\[
W(d) \;=\; \sum_{l=1}^L \bigl(n_{l-1}n_l+n_l\bigr).
\]

\paragraph{Direct Generator.}
The direct model outputs $2d^2$ real parameters (corresponding to the real and imaginary 
entries of $\rho$). 
Generation cost is therefore dominated by the MLP forward pass:
\[
\mathcal{T}_{\text{Direct}}(K) \;=\; \Theta\!\bigl(K W(d)\bigr).
\]
Because the final affine map must produce $2d^2$ outputs,
\[
W(d) \;\ge\; \Omega(d^2).
\]
Thus, $\mathcal{T}_{\text{Direct}}(K) = \Omega(K d^2)$ always holds. 
If hidden widths scale sub-cubically, $W(d)=o(d^3)$, the direct model avoids any $\Omega(d^3)$ 
term at inference.\\

\paragraph{Decomposition Generators.}
The decomposition models output $O(d^2)$ parameters that define a factorization 
(e.g., Cholesky or LDL). 
For each state, the following steps are required:\\
1. MLP forward pass: $\Theta(W(d))$,\\
2. One dense $d\times d$ complex matrix multiplication to form $\rho$: $\Theta(d^3)$,\\
3.Trace normalization and any additional assembly: $o(d^3)$.\\

The per-state complexity is therefore
\[
\mathcal{T}_{\text{Decomp}}(K) \;=\; \Theta\!\bigl(K(W(d)+d^3)\bigr).
\]
In particular, $\mathcal{T}_{\text{Decomp}}(K) \;\ge\; \Omega(K d^3)$, 
as at least one $d\times d$ multiplication is unavoidable.\\

\paragraph{Implications.}
Both approaches scale linearly in $K$, but their dependence on $d$ differs:\\
1. Direct: $\Theta(KW(d))$ with $W(d)\ge \Omega(d^2)$, which is asymptotically more efficient in $d$ whenever $W(d)=o(d^3)$.\\
2. Decomposition: $\Theta(K(W(d)+d^3))$, with the $d^3$ term dominating unless the network width already grows at least cubically.\\
As a concrete example, at $d=8$ one dense multiplication requires $d^3=512$ complex multiplications, 
or approximately $4096$ real floating-point operations. 
Decomposition models guarantee physicality by construction but incur this $\Theta(d^3)$ overhead, 
whereas the direct model trades physical guarantees for potentially lower inference cost.\\

\paragraph{Empirical validation.}
To corroborate these asymptotic estimates, we performed machine-agnostic micro-benchmarks 
using synthetic generators without training. 
We measured per-state wall-clock time for forward passes, decomposition assembly, 
and validation (PSD and PPT checks) across a broad range of Hilbert space dimensions 
$d \in \{4,6,8,12,\ldots,256\}$ and batch size $B=1024$, 
on both CPU and GPU backends. 
Figure~\ref{fig:scaling} summarizes the results. 
Forward-pass costs (Direct, Cholesky, LDL) scale nearly quadratically with $d$, 
consistent with the $\Theta(d^2)$ dependence on parameter count. 
By contrast, decomposition assembly ($LL^\dagger$, $LDL^\dagger$) 
and eigenvalue-based PSD/PPT checks exhibit cubic growth $\Theta(d^3)$, 
as predicted by the analytic model. 
The same trends hold on CPU and GPU, confirming that the complexity analysis 
is robust and machine-agnostic. 
Error bars denote 95\% confidence intervals across 10 repeats. \\

\paragraph{Benchmark configuration.}
All scalability measurements were performed on both CPU and GPU backends using 
synthetic generators without training. 
We evaluated Hilbert space dimensions 
$d \in \{4, 8, 12, 16, 20, 24, 28, 32, 36, 40, 64, 96, 128, 160, 256, 512\}$ 
and batch sizes 
$B \in \{64, 128, 256, 512, 1024\}$. 
Each measurement was averaged over 15 independent repeats, with 95\% confidence intervals 
reported as error bars. 
CPU experiments were restricted to four threads for consistency, 
and PPT checks were omitted beyond $d=32$ to prevent prohibitive cubic overhead. 
Since models were not trained, these benchmarks isolate the inference and validation cost of forward passes, decomposition assembly, and Positive Semi Definite checks.

\section{Mathematical Derivations}
\label{app:derivations}

\subsection{Equivalence of Entanglement and Teleportation fidelity for Bell-Diagonal States}

\emph{Definition.} Bell-diagonal states have the density matrix form:
\begin{equation}
\rho_{\text{Bell}} = \frac{1}{4}\left(\mathbb{I} \otimes \mathbb{I} + \sum_{i=1}^{3} c_i \sigma_i \otimes \sigma_i\right),
\end{equation}
with parameters satisfying $-1 \leq c_i \leq 1$, and $\sigma_i$ being the Pauli matrices.

\emph{Entanglement criterion (Peres-Horodecki).} A Bell-diagonal state is entangled if and only if its partial transpose has at least one negative eigenvalue. For this class of states, this is equivalent to:
\begin{equation}
|c_1| + |c_2| + |c_3| > 1.
\end{equation}

\emph{Teleportation fidelity criterion (Horodecki).} The maximal teleportation fidelity $F_{\text{max}}(\rho)$ for Bell-diagonal states is given by:
\begin{equation}
F_{\text{max}}(\rho) = \frac{1}{2}\left(1 + \frac{|c_1| + |c_2| + |c_3|}{3}\right).
\end{equation}
A quantum advantage (i.e., the state is a useful resource for teleportation) is achieved when $F_{\text{max}}(\rho) > \frac{2}{3}$.

\emph{Proof of equivalence.} Substituting the fidelity condition and simplifying, we obtain:
\begin{align}
\frac{1}{2}\left(1 + \frac{|c_1| + |c_2| + |c_3|}{3}\right) &> \frac{2}{3} \\
1 + \frac{|c_1| + |c_2| + |c_3|}{3} &> \frac{4}{3} \\
\frac{|c_1| + |c_2| + |c_3|}{3} &> \frac{1}{3} \\
|c_1| + |c_2| + |c_3| &> 1.
\end{align}
\subsection{Equivalence of Teleportation fidelity and PPT Inseparability for Werner-like Two-Qubit States}
\label{app:teleportation_ppt_equivalence}

\noindent For the Werner-like family of two-qubit states considered in this work, the condition for teleportation fidelity coincides exactly with the PPT inseparability condition.

\noindent \paragraph{State definition.}
Consider the two-qubit state
\begin{equation}
\rho(p,\alpha)
= p\,|\psi\rangle\langle\psi|
+ \frac{1-p}{4}\,I_4,
\qquad 0 \le p \le 1,
\end{equation}
where
\begin{equation}
|\psi\rangle = \alpha |00\rangle + \beta |11\rangle,
\qquad
\alpha,\beta \ge 0,
\qquad
\alpha^2+\beta^2=1.
\end{equation}
In Bloch form, the diagonal correlation tensor is
\begin{equation}
T = \mathrm{diag}\!\left(2p\alpha\beta,\,-2p\alpha\beta,\,p\right).
\end{equation}

\paragraph{Teleportation Fidelity.}
For a two-qubit state with diagonal correlation tensor $T$, the maximal teleportation fidelity is
\begin{equation}
F_{\max}(\rho)
= \frac12\left(1+\frac13 N(\rho)\right),
\end{equation}
\begin{equation}
N(\rho)=\sum_{i=1}^3 |t_i| = p(1+4\alpha\beta).
\end{equation}
Teleportation is useful if $F_{\max}>2/3$, equivalently $N(\rho)>1$, yielding
\begin{equation}
\label{eq:teleportation_condition}
p(1+4\alpha\beta)>1.
\end{equation}

\paragraph{PPT inseparability.}
The partial transpose $\rho^{T_B}$ has smallest eigenvalue
\begin{equation}
\lambda_-=\frac{1-p}{4}-p\alpha\beta.
\end{equation}
The state is entangled (PPT-negative) iff $\lambda_-<0$, which reduces to
\begin{equation}
\label{eq:ppt_condition}
p(1+4\alpha\beta)>1.
\end{equation}

\paragraph{Equivalence.}
Comparing Eqs.~\eqref{eq:teleportation_condition} and~\eqref{eq:ppt_condition}, teleportation-useful and PPT-negative conditions coincide for this family. Both are governed by the boundary
\begin{equation}
p=\frac{1}{1+4\alpha\beta}
=\frac{1}{1+4\sqrt{\alpha^2(1-\alpha^2)}}.
\end{equation}
The commonly cited threshold $p>1/3$ corresponds to maximal entanglement ($\alpha^2=\beta^2=1/2$) and is necessary but not sufficient when $\alpha$ is unrestricted. This equivalence is specific to the Werner-like two-qubit family.

\end{document}